\documentclass[english,aps,pra,reprint,superscriptaddress,longbibliography]{revtex4-2}
\usepackage[caption=false]{subfig}
\usepackage[T1]{fontenc}
\usepackage[latin9]{inputenc}
\usepackage{amssymb}
\usepackage{graphicx,xcolor,array, colortbl}
\usepackage{subscript}
\usepackage{amsmath}
\usepackage[colorlinks=true,citecolor=blue,urlcolor=blue,linkcolor=blue]{hyperref}

\makeatletter
\usepackage{pslatex}

\usepackage{hyperref}
\usepackage{xcolor}
\hypersetup{
    colorlinks,
    linkcolor={blue!70!black},
    citecolor={blue!70!black},
    urlcolor={blue!70!black}
}

\usepackage{babel}
\usepackage{comment}

\newcommand{\X}{\tilde{X}^{2}\Sigma^{+}}
\newcommand{\A}{\tilde{A}^{2}\Pi_{1/2}}
\newcommand{\B}{\tilde{B}^{2}\Sigma^{+}}
\newcommand{\wn}{~\rm{cm}^{-1}}

\newcommand{\MOTbaseline}{7200} 
\newcommand{\MOTbaselineUnc}{1400} 
\newcommand{\MOTnum}{32400} 
\newcommand{\MOTnumUnc}{4700} 
\newcommand{\numlifetime}{66} 
\newcommand{\numlifetimeunc}{5.6} 

\newcommand{\fullbudget}{15000}
\newcommand{\fullbudgetunc}{1400}
\newcommand{\tenlasersbudget}{9800}
\newcommand{\tenlasersbudgetunc}{1000}

\newcommand{\quarterVtenLlifetime}{99}
\newcommand{\quarterVtenLlifetimeUNC}{7.6}

\newcommand{\fullbudgetloadingtau}{14}
\newcommand{\fullbudgetloadingtauunc}{1.4}

\newcommand{\maxlifetime}{210}
\newcommand{\maxlifetimeUnc}{35} 
\newcommand{\maxlifetimeten}{99}
\newcommand{\maxlifetimetenUnc}{7.6}

\newcommand{\oneVcal}{6.0}
\newcommand{\lowPowerHold}{1} 

\newcommand{\ellten}{$1.0(13)\times10^{-4}$}
\newcommand{\elltwelve}{$3.3(20)\times10^{-5}$}

\newcommand{\XBrepumpcm}{15931.49 cm$^{-1}$}

\newcommand{\SRsplitting}{106.7} 

\newcommand{\paritydoubletsplitting}{280} 

\newcommand{\naturallinewidth}{7} 

\newcommand{\Bvalpaper}{17148.577(2)} 

\begin{document}
\title{High-sensitivity molecular spectroscopy of SrOH using magneto-optical trapping}

\author{Annika Lunstad}
\email{alunstad@g.harvard.edu}
\affiliation{Harvard-MIT Center for Ultracold Atoms, Cambridge, Massachusetts 02138, USA}
\affiliation{Department of Physics, Harvard University, Cambridge, Massachusetts 02138, USA}

\author{Hiromitsu Sawaoka}
\altaffiliation{Current address: Department of Physics, University of California, Berkeley, California 94720, USA}
\affiliation{Harvard-MIT Center for Ultracold Atoms, Cambridge, Massachusetts 02138, USA}
\affiliation{Department of Physics, Harvard University, Cambridge, Massachusetts 02138, USA}

\author{Zack Lasner}
\altaffiliation{Current address: IonQ, Inc., College Park, Maryland 20740, USA}
\affiliation{Harvard-MIT Center for Ultracold Atoms, Cambridge, Massachusetts 02138, USA}
\affiliation{Department of Physics, Harvard University, Cambridge, Massachusetts 02138, USA}

\author{Abdullah Nasir}
\affiliation{Harvard-MIT Center for Ultracold Atoms, Cambridge, Massachusetts 02138, USA}
\affiliation{Department of Physics, Harvard University, Cambridge, Massachusetts 02138, USA}

\author{Mingda Li}
\affiliation{Harvard-MIT Center for Ultracold Atoms, Cambridge, Massachusetts 02138, USA}
\affiliation{Department of Physics, Harvard University, Cambridge, Massachusetts 02138, USA}

\author{Jack Mango}
\affiliation{Harvard-MIT Center for Ultracold Atoms, Cambridge, Massachusetts 02138, USA}
\affiliation{Department of Physics, Harvard University, Cambridge, Massachusetts 02138, USA}

\author{Rachel Fields}
\affiliation{Harvard-MIT Center for Ultracold Atoms, Cambridge, Massachusetts 02138, USA}
\affiliation{Department of Physics, Harvard University, Cambridge, Massachusetts 02138, USA}

\author{John M. Doyle}
\affiliation{Harvard-MIT Center for Ultracold Atoms, Cambridge, Massachusetts 02138, USA}
\affiliation{Department of Physics, Harvard University, Cambridge, Massachusetts 02138, USA}

\begin{abstract}
    
    Polyatomic molecules are projected to be powerful tools in searches for physics beyond the Standard Model (BSM), including new CP-violating (CPV) interactions and ultralight dark matter (UDM) particles~\cite{kozyryev2021enhanced,kozyryev2017precision}. Certain degrees of freedom present in polyatomic molecules enhance the sensitivity of these searches,  as well as reject systematic errors, but necessitate extensive high-precision spectroscopy to identify pathways for optical cycling and quantum state readout. Here we show how a magneto-optical trap (MOT) can be used to locate weak optical  transitions and identify rovibronic states for optical cycling and quantum control. We demonstrate this spectroscopic approach with strontium monohydroxide (SrOH), which is a candidate for both CPV and UDM searches ~\cite{LasnerMOT,kozyryev2021enhanced,kozyryev2017precision}. We identify two new repumping transitions in SrOH and implement them in a deeper optical cycle to achieve \MOTnum(\MOTnumUnc) trapped molecules, a 4.5-fold increase over the previous, shallower cycle. In addition, we determine the energy spacing between the $\X(200)$ and $\X(03^10)$ vibrational manifolds of SrOH, confirming the existence of numerous low-frequency rovibrational transitions that are sensitive to temporal variations of the proton-to-electron mass ratio, a predicted effect of the existence of UDM.
\end{abstract}

\maketitle

\section{Introduction}

Molecules offer rich structures with several degrees of freedom that cover a wide range of energy scales. Despite the complexity of even small molecules, much has already been achieved, including simultaneous quantum control of electronic, vibronic, and rotational degrees of freedom, and laser cooling~\cite{Anderegg2019,Vilas2024}. Driven by potential applications in quantum computing~\cite{Demille2002,Ni2018,Sawant2020}, quantum simulations~\cite{Micheli2006, Gorshkov2011}, and precision searches for physics beyond the Standard Model (BSM), a large variety of ultracold molecules, both laser cooled and assembled, are currently under study~\cite{Kobayashi2019,Zeng2024,Christakis2023,Holland2025,Alauze2021,Picard2025,Miller2024,Burau2023,Gaul2024}. 

In the area of searches for BSM physics, cold diatomic molecules have set the strictest bounds on the CP-violating (CPV) electron electric dipole moment (eEDM)~\cite{JILAedm,ACMEII}, broadly constraining new CPV physics in the 1--10~TeV range. Heavy polyatomic molecules are new candidates for both electronic and nuclear CPV searches~\cite{kozyryev2017precision,Augenbraun2020,Hao2020,Gaul2024,Isaev2010}. 
Molecules are also sensitive probes of dark matter. The spectra of molecules will shift due to temporal variations of the proton-to-electron mass ratio, $\mu\equiv m_p/m_e$, through differential shifts between rovibronic levels \cite{Kobayashi2019,Hanneke2021}. In polyatomic molecules, the different degrees of freedom of vibration (e.g., bend and stretch) generically allow for near-degenerate vibrational states, offering accessible transitions sensitive to $\mu$ in the convenient microwave regime. Probes of such shifts provide a direct window onto many models of bosonic ultralight dark matter (UDM)~\cite{Ferreira2020}. 
The molecule we study here, strontium monohydroxide (SrOH), has sensitivity to both eEDM and UDM~\cite{kozyryev2021enhanced,kozyryev2017precision}.

Control over single quantum states and the trapping of large numbers of molecules are desirable to achieve optimal conditions for precision measurements and other quantum science applications. 
Ultracold, trapped neutral molecules have been produced via opto-electric Sisyphus cooling~\cite{Prehn2016}, association of ultracold atoms~\cite{Liu2018},  and direct laser cooling~\cite{Langin2021, LasnerMOT}. 
One key distinction of laser cooling in molecules, as opposed to atoms, is the existence of radiative decays to vibrationally excited states. Molecules populating these states must be returned (i.e., repumped) to the vibrational ground state to sustain an optical cycle and continue the cooling process. 
By addressing ten rovibrational states with lasers, a magneto-optical trap (MOT) of SrOH was recently achieved~\cite{LasnerMOT}. In that work, a large fraction of molecules still populated unaddressed vibrational states over the course of cooling and trapping, motivating further spectroscopy to identify additional repumping transitions. Driving new repumping pathways would increase the number of trapped molecules and the concomitant sensitivity of any precision measurement.

In this work, we develop the MOT as a spectroscopy resource to probe weak transitions and present methods to fully identify discovered vibronic levels. Using this MOT-based spectroscopy, we report two new repumping transitions for the previously unaddressed $\X(12^00)$ and $\X(12^20)$ states in SrOH and implement them to realize improved laser cooling and trapping. The addition of these repumps provides a 4.5-fold increase in the number $N$ of SrOH molecules trapped in a MOT. Combining this increase with additional technical improvements, we realize $N=$\MOTnum(\MOTnumUnc), an order of magnitude improvement over previous work. This, in turn, recently enabled $>$$10^3$ SrOH molecules to be captured in an optical dipole trap (ODT)~\cite{Sawaoka2025}. Also, for the first time, we locate the vibrational manifolds that are key to proposed measurements of UDM with SrOH~\cite{kozyryev2021enhanced} and demonstrate a pathway to directly populate it. In doing so, we identify many low-frequency (5--100~GHz) rovibrational transitions with sensitivity to small variations in $\mu$, thus validating precision SrOH spectroscopy as a platform for near-future probes of UDM ~\cite{kozyryev2021enhanced}.

\section{Experimental method}\label{sec:methods}

A MOT offers long interaction times and---through the many repump pathways used to laser cool---easy access to numerous excited vibrational states within the ground electronic state manifold. Taking advantage of rapid optical cycling, MOT fluorescence measurements have the potential to be a very useful tool for identifying weak transitions and finding excited vibrational states, in both ground and excited electronic manifolds. Here, we develop the MOT as a vibrational spectroscopy tool to identify new, very weak optical cycling transitions. Our apparatus and laser cooling procedure was described in~\cite{LasnerMOT}, with the few minor changes to the optical cycle for this work described in the Appendix~\ref{SI:cyclingscheme}. 
We first capture molecules in the MOT from a laser-slowed cryogenic buffer gas beam~\cite{LasnerMOT}. After initial capture, the molecules cool in the MOT over a time period of $\sim$20~ms to a temperature of $\sim$1~mK. A molecule can decay out of the optical cycle, such as to a dark vibrational state that is not repumped; it is then no longer confined by the MOT forces. However, due to its finite velocity, this molecule will remain in the spatial region of the MOT for  $\sim$10~ms and, if repumped out of its dark state, can be recaptured in the MOT and returned to the optical cycle.
This free flight time allows for sensitive "repumper spectroscopy," as described below. 

\subsection{Repumper spectroscopy for ground states}

Consider a transition of interest as a two-level system with no loss from the ground state and finite loss from the excited state, to outside of the two-level system. During ``repumper spectroscopy,'' this ``loss'' from the two-level system is into the optical cycle and manifests itself as increased MOT fluorescence. In this repumper spectroscopy, we start in the situation where the energy of the excited state is known and we seek to accurately determine the energy of the ground state, referred to as the ``target state.'' 
From Appendix~\ref{App:lineshape}, the population removed from this two-level system after time $t$, in the large detuning limit, is $$N(t)\propto1-\exp\left[-\frac{\gamma}{2}\frac{\Omega^2}{\Delta^2}t\right],\label{eq:numremoved}$$ where $\gamma$ is the decay rate from the excited state (to a state outside of the two level system under consideration), $\Omega$ is the Rabi frequency, and $\Delta$ is the detuning. The characteristic removal time is $\tau_\text{r} =2\Delta^2/(\gamma~\Omega^2)$. Defining the saturation parameter $s\equiv2(\Omega/\gamma)^2$, typical values for spectroscopy with SrOH in the MOT are $s=1$, $\gamma=2\pi\times7$~MHz, and $\tau_\text{r}=10$~ms, yielding a characteristic detuning of $\Delta =2\pi\times 2.3$~GHz. Thus, at a detuning about $10^3$ times larger than the natural linewidth, optical pumping out of the target state can be observed because decay out of the two-level system under consideration populates the optical cycle and manifests itself as increased MOT fluorescence.
From a practical perspective, this sensitivity at very large detunings is a feature and a huge advantage of repumper spectroscopy in the MOT, enabling much quicker searches compared to measurements in molecular beams with short interaction times.

Throughout, we employ the notation $(v_1 v_2^{\ell}v_3)$ denoting the vibrational state with $v_1$ quanta of excitation in the Sr-O stretching mode, $v_2$ quanta in the Sr-O-H bending mode with vibrational orbital angular momentum $\ell\hbar$, and $v_3$ quanta in the O-H stretching mode. To accurately determine the energy of the target state, we seek to drive a transition from it back into the optical cycle. However, the target state first needs to be populated, which we accomplish in one of two ways using the MOT. (1) The molecules may be directly pumped into the target state, for example driving $\X(010)$--$\tilde{A}(030)\kappa^2\Sigma_{1/2}$, which decays predominantly to $\X(03^10)$. (2) Alternatively, optical cycling may indirectly populate a state through $\Delta v_1=1$ decays due to off-diagonal branching. For example, driving $\X(02^00)$--$\tilde{A}(020)\mu^2\Pi_{1/2}$  populates $\X(12^00)$, beyond the already existing population from decays out of the optical cycle.

Starting with molecules in the target state, we seek to repump them through a known excited state. We scan a repumping laser frequency to find the target state by recovering the MOT. In our case, the MOT cloud of SrOH is addressed using a separate repumping spectroscopy light beam~\footnote{For the repumping transitions driven in this work, the light was produced by a Ti:Sapph laser with a wavelengths of $\sim$700-712~nm and intensity of $I\sim900$~mW/cm$^2$.}. The frequency of this light is scanned to search for MOT recovery and so acts as a repump laser. The number of molecules in the MOT is measured in the following three conditions for each repump laser frequency: 1) baseline MOT number, $N_1$, 2) the MOT number after depletion into the target state, $N_2$, and 3) the MOT number after depletion and recovery via the repump laser, $N_3$. We show the ratio of the recovered MOT population to the depleted MOT population, $(N_3-N_2)/(N_1-N_2)$, for an example scan in Figure~\ref{fig:peaks}(b). Signal from repumper transitions is observed at detunings as large as $\Delta \sim$ $2\pi\times6$~GHz. 

Once a repumping transition is found, we confirm its origin by measuring MOT fluorescence in all four combinations of depletion on or off, and repumping on or off. This determines whether the transition recovers the population that had been depleted from the MOT to a target state, or whether the transition improves the MOT fluorescence by adding molecules to the cycle from some other state. Additionally, finer scans with lower laser intensity, as in Figure~\ref{fig:peaks}(c) can then provide frequency resolution at the level of the natural linewidth, $\gamma\sim2\pi\times10$~MHz. 

\subsection{Depletion spectroscopy for excited states}
The long interaction times in the MOT lead to signal at large detunings when finding a repumping line (above) and also allow weak lines to be driven with a high signal-to-noise ratio, which is useful when searching for excited states. 

\begin{figure}[] 
    \centering
    \includegraphics[width=\columnwidth]{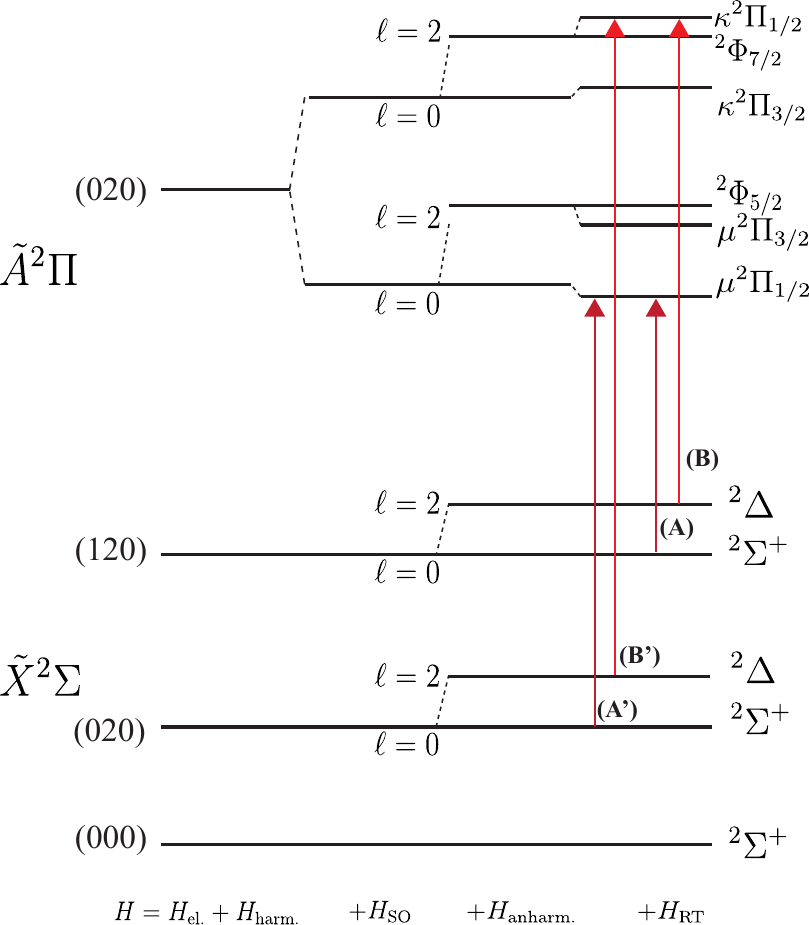} 
    \caption{Energy levels used for spectroscopy as well as repumping transitions (not to scale; shifts below $\sim$1~$\wn$ not shown). The splittings are ordered by their size, which can be estimated from the relevant constants: $A = 263.5$~$\wn$, $\ell^2g_{22}=4\times7.656$~$\wn$, and $(\epsilon \omega_2)^2/A=(-30.72)^2/263.5=3.6$~$\wn$~\cite{SrOHA010}. Lines labeled case (A) and (B) show the transitions relevant for $\X(12^00)$ and $\X(12^20)$ repumper spectroscopy, respectively.}
    \label{fig:energylevels}
\end{figure}

For excited-state spectroscopy, a narrowband laser is directed at the MOT and its frequency is scanned. When the laser light frequency approaches resonance, the molecules are driven to excited states, which, for the cases of interest here, predominantly decay to optically {\sl{unaddressed states}}. Thus, molecules are driven out of the optical cycle, resulting in a depletion of MOT fluorescence, as in Figure~\ref{fig:peaks}(a). We refer to this method as ``depletion spectroscopy.'' For excited state searches described here with SrOH, the beam intensity is $I\sim $3.5~mW/cm$^2$ and the wavelength is $\sim$640~nm \footnote{The light is generated by a continuous-wave 899 Coherent Ring dye laser with Rhodamine 640 as the gain medium pumped by 10~W of 532~nm light. When populating states $\X(12^{0,2}0)$ via cycling on a diagonal line to deplete (transitions (A) and (A') in Figure~\ref{fig:energylevels}), homebuilt ECDLs were used.}.

\begin{table}
    \centering
    \begin{tabular}{|c|c|c|c|c|c|}\hline
 $\Lambda$& $\ell$& $\Sigma$& $K=\Lambda+\ell$& $P=\Lambda+\ell+\Sigma$&Term symbol\\\hline \hline
         +1&  +2&  1/2&  3&  7/2& $^2\Phi_{7/2}$\\\hline
         +1&  0&  1/2&  1&  3/2& $\kappa^2\Pi_{3/2}$\\\hline
 +1& -2& 1/2& -1& -1/2&$\kappa^2\Pi_{1/2}$\\\hline
         -1&  +2&  1/2&  1&  3/2& $\mu^2\Pi_{3/2}$\\\hline
         -1&  0&  1/2&  -1&  -1/2& $\mu^2\Pi_{1/2}$\\ \hline
 -1& -2& 1/2& -3& -5/2&$^2\Phi_{5/2}$\\ \hline
    \end{tabular}
    \caption{Term symbols and approximate angular momentum character of states studied in this work. In the presence of the Renner-Teller interaction, only $K$ and $P$ are good quantum numbers. In SrOH, states predominantly have the angular momentum composition given above because the Renner-Teller parameter is small compared to the spin-orbit constant.}
    \label{tab:Astates}
\end{table}

Because of the many vibrational and rotational states within a molecule, once a transition is observed, further measurements are required to fully identify the ground and excited states involved. For example, consider a depletion feature in the MOT fluorescence at some frequency, corresponding to a transition $x_g - x_e,$ where $x_g~(x_e) $  is a ground (excited) vibronic manifold. For our cycling scheme in SrOH, any observed state in the vibronic manifold $x_g$ can have only $J=1/2$ and $J=3/2$ rotational levels populated, where $J$ is the total angular momentum excluding nuclear spin, because all addressed excited states in the optical cycle have $J=1/2$. As a result, molecules can be driven through rotational states up to $J=5/2$ in the excited state, due to angular momentum selection rules. Identifying each rotational feature and the rotational state splittings provides information for assigning $x_e$, since the lowest possible rotational level for the excited vibronic manifolds studied in this work range from $J=1/2$ to $J=3/2$. Additionally, in any excited manifold suitable for repumping, it is necessary to find the $J=1/2$ state in order to retain rotational closure.

The next step in confirming the assignment of $x_e$ is identifying $x_g$, which is done using a depletion-based assignment procedure as follows. After capturing molecules in the MOT, light addressing one candidate ground state $x_i$ is turned off for 3~ms during optical cycling, where $i=0,\dots,N$, and $i$ labels a laser frequency employed in the optical cycle, and $N=10$ in our experiment. This pumps a significant fraction of molecules into $x_i$. Next, only the depletion light is turned on (for 10~ms in our experiment), resulting in depletion of molecules out of $x_i$. Finally, the molecules are recaptured in the MOT with all repumps on. 
If $x_i=x_g$, the depletion light pumps molecules out of the optical cycle, lowering the recaptured MOT fluorescence. For the other states $x_j$ where $x_j\neq x_g$, the recaptured MOT fluorescence is unaffected. This procedure determines the ground state of a given depletion feature. The excited state rotational structure, laser frequency, and ground state assignment are typically sufficient to identify the excited state in the transition. We sometimes also confirm the state assignment by observing the decay fluorescence frequency on a spectrometer, as described in Appendix~\ref{SI:otherstates}. 

This depletion spectroscopy method applied to SrOH is sufficient to fully identify and label the states for the $\tilde{A}(020)$ system by driving from $\X(000)$ to one of the $\tilde{A}(020)$ states. Specifically, the $\A(020)\mu^2\Pi_{1/2}$ and $\A(020)\kappa^2\Pi_{1/2}$ states are used for repumping $\X(12^00)$ and $\X(12^20),$ respectively. The $\tilde{A}(020)\mu^2\Pi_{1/2}$ state had already been identified for previous work~\cite{LasnerMOT}. 
Furthermore, the excited state $\tilde{A}(030)\kappa^2\Sigma_{1/2}$, mainly of $\ell=1$ character, was identified with these methods using the transition $\X(010) - \tilde{A}(030)\kappa^2\Sigma_{1/2}$. Several other excited states were identified and are listed in the Appendix~\ref{SI:otherstates}. By performing the excited state spectroscopy using depletion of the MOT, we achieve high-contrast signal even for weak transitions such as those with $\Delta v_2=2.$

Using the methods described above, we find $\X(12^00),~\X(12^20),$ and $\X(03^10)$, as well as the excited states $\tilde{A}(020)$ with $\Pi$ vibronic symmetry and $\tilde{A}(030)\kappa^2\Sigma_{1/2}$. With the precise knowledge of the $\X(03^10)$ state energy from this work, we determine that many energy spacings between low-lying rotational levels in the $\X(200)$--$\X(03^10)$ band lie in the 1--100~GHz range, confirming the viability of a proposed UDM search using temporal variations of $\mu$~\cite{kozyryev2021enhanced}, as discussed further in Appendix~\ref{App:sciencestates}.

\section{Description of Spectroscopy results}

\begin{figure}[]
    \centering
    \includegraphics[width=.95\columnwidth]{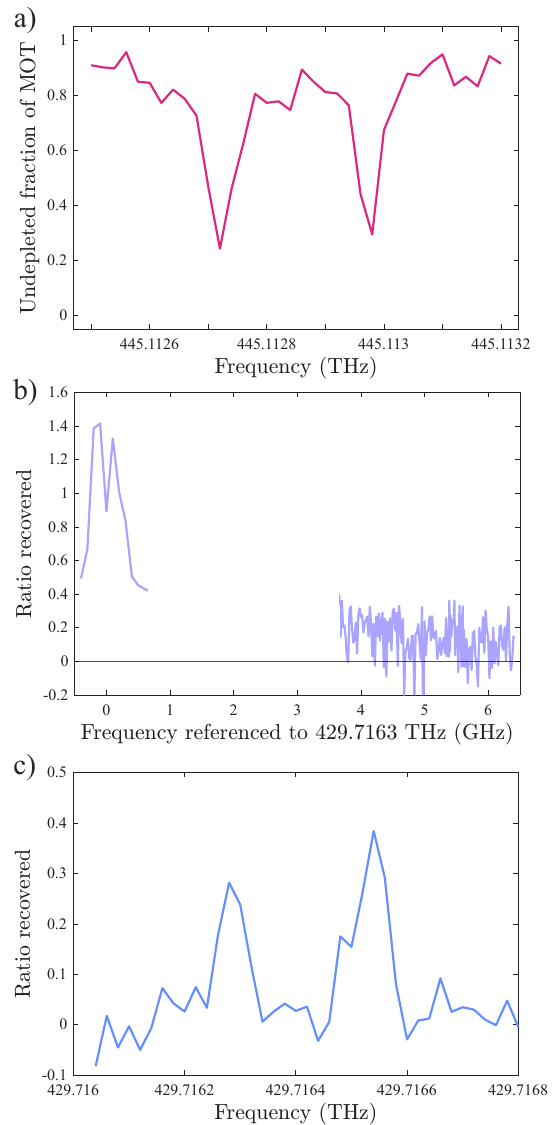}
    \caption{Spectrosopy data (a) scan over $\X(02^20)$--$\tilde{A}(020; J=1/2)\kappa^2\Pi_{1/2}$, showing both excited state parity components, split by $\sim$\paritydoubletsplitting~MHz. (b) repumper spectroscopy of $\X(12^20)$--$\tilde{A}(020)\kappa^2\Pi_{1/2}$ at low resolution. 
    Measurable repumping is observed over 4~GHz detuned from the transition, as expected from the long spectroscopic interaction times achieved in the MOT, see Sec.~\ref{sec:methods}. 
    (c) a finer scan of the same transition with 0.1\% of the full power in the repumping light, revealing the parity doublet splitting in the excited state.}
    \label{fig:peaks}
\end{figure}

Before we began our work with SrOH, it had been studied in the spectroscopic literature by several authors over the course of more than a decade~\cite{SrOH020, SrOHA010,Fletcher95,Presunka95,Brazier85,Nakagawa83}. These earlier results provide an essential background for predicting the transitions and estimating the splittings in this work. The $\tilde{A}^2\Pi$ state in SrOH is a Hund's case (a) state with molecule-frame electronic orbital angular momentum projection $\Lambda = \pm1$ and electron spin projection $\Sigma=\pm1/2$. The spin-orbit interaction couples $\Lambda$ and $\Sigma$ to split the $\tilde{A}^2\Pi$ state into $\A$ and $\tilde{A}^2\Pi_{3/2}$, which are separated by $263~\wn$. In states with $|\ell|>0$, the electronic and vibrational angular momenta combine to form $K=\Lambda+\ell$, which determines the vibronic symmetry of the state. The molecule-frame projection of total angular momentum excluding nuclear spin is $P = \Lambda + \ell + \Sigma$. A state's character is designated by the term symbol $^{2 S+1}|K|_{|P|}$, where $|K|$ is labeled as $\{\Sigma,\Pi,\Delta,\dots \}$ for values of $|K|=\{0, 1, 2,\dots\}$. States with the same term symbols are distinguished by $\mu$ and $\kappa$ to designate lower and higher energies, respectively, as seen in Figure~\ref{fig:energylevels}. The term symbols relevant for the $\tilde{A}(020)$ states are additionally written explicitly in Table~\ref{tab:Astates}, and we use the vibronic notation described here throughout. It is important to note that in the $\tilde{A}^2\Pi$ state, the actual eigenstates of the molecule are a mixture of $\ell=0 \text{ and }2$ due to Renner-Teller interactions with strength governed by the parameter $\epsilon=-0.0791$~\cite{SrOHA010}. Thus, for example, the $\tilde{A}(020)\mu^2\Pi_{1/2}$ state is primarily $\ell=0$ character but is mixed with $\ell=2$ at the $\sim2\%$ level (see Figure~\ref{fig:energylevels}).

When searching for excited states, the long interaction times in the MOT enable near-complete depletion even for weak transitions. By either driving directly from the ground $\X(000)$ state or repumping one of the $\X(120)$ states, the $\tilde{A}^2\Pi (020) $ states of $\mu^2\Pi_{1/2}, \mu^2\Pi_{3/2}, \kappa^2\Pi_{3/2},$ and $\kappa^2\Pi_{1/2}$ vibronic symmetry were identified and are listed in Table~\ref{tab:energies}. The vibronic $^2\Phi_{5/2}$ and $^2\Phi_{7/2}$ states were not observed.

When searching for ground states, long interaction times likewise enable large step sizes of 100s of MHz to observe initial signal, with non-negligible repumping signal up to $\sim6$~GHz from the nearest resonance as shown in Figure~\ref{fig:peaks}(b). This effect is $\sim$1,000 times broader than the molecule's natural linewidth of $2\pi\times$\naturallinewidth~MHz. This is an important, defining feature of our technique. Subsequently, employing smaller step sizes ($\sim$10~MHz) while lowering the intensity of the repumping light allows for finer mapping of the line shape as shown in Figure~\ref{fig:peaks}(c). With this finer mapping, we identify the $\X(120)$ states to $\sim$10~MHz resolution, with all transitions listed in Table~\ref{tab:alltransitions}.

In addition to the most relevant states already listed, we found several other states consistent with $\Delta v_2 =+2$ bands. In each case, excited states were later identified with dispersed laser-induced fluorescence, as described in Appendix~\ref{SI:otherstates}. 

\section{Characterization of MOT improvement}

With the new knowledge obtained from both repumper and depletion spectroscopy, we add the repumping laser frequencies addressing the $\X(12^00)$--$\tilde{A}(020)\mu^2\Pi_{1/2}$ and $\X(12^20)$--$\tilde{A}(020)\kappa^2\Pi_{1/2}$ transitions to the optical cycling scheme during the laser slowing and magneto-optical trapping. We observe more efficient trapping due to the improved optical cycle. Both new transitions are driven by ECDLs with sufficient power to not limit the overall photon scattering rate during slowing. As with other repumpers, the light is broadened to $\sim$350~MHz by an over-driven electro-optic modulator. Adding these transitions extends the average number of photons cycled before loss to an unaddressed state, which in turn increases both the maximum trapped molecule number and the MOT lifetime. With these additional two light frequencies, the highest observed MOT lifetime increased from \quarterVtenLlifetime(\quarterVtenLlifetimeUNC)~ms to \maxlifetime(\maxlifetimeUnc)~ms as shown in Figure~\ref{fig:lifetimes_powers}. Additionally, the loss probability in the optical cycle decreased from \ellten\ to \elltwelve, as discussed in Appendix~\ref{App:number}. As a result, the MOT number increased by a factor of 4.5 from $N=\MOTbaseline(\MOTbaselineUnc)$ to $N=\MOTnum(\MOTnumUnc)$, as a direct consequence of the addition of these repumpers. 

\begin{figure}
    \centering
    \includegraphics[width=\columnwidth]{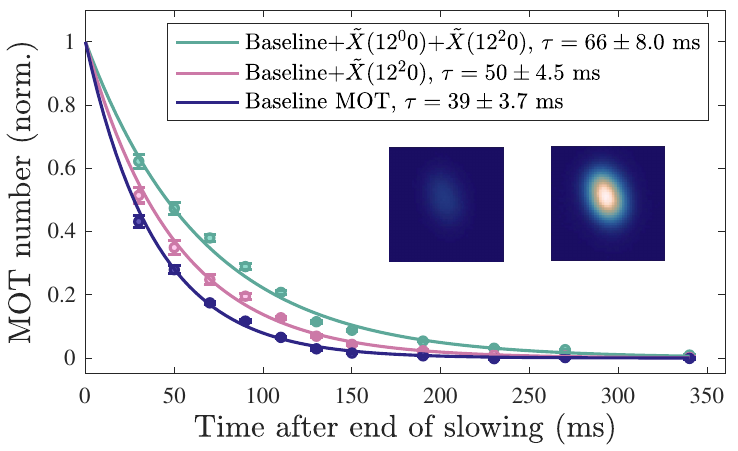}
\caption{Increase in MOT lifetime with sequentially added repumping lasers at a ``hold'' MOT laser power of \oneVcal~mW. The data and fit for adding only the $\X(12^00)$ repumping laser is omitted for visual clarity. The baseline MOT includes all ten lasers in the original optical cycle, described in~\cite{LasnerMOT}. For conditions:\{Baseline, Baseline + $\X(12^00)$, Baseline + $\X(12^20)$, Baseline + $\X(12^00)$ + $\X(12^20)$\}, the lifetimes are $\{39.4(3.7),~49.4(3.9),~50.4(4.5),~66.3(8.0)\}$~ms, respectively. Images show fluorescence from trapped molecule clouds with peak molecule numbers of \MOTbaseline(\MOTbaselineUnc) and \MOTnum(\MOTnumUnc), corresponding to optical cycles without (left) and with (right) the $\X(12^{\ell}0)$ repumping lasers. The same color scale is used in both images.}
    \label{fig:lifetimes}
\end{figure}

Several mechanisms contribute to improving the number of trapped molecules. The most significant contributor is the direct increase of the average photon scatters before $1/e$ decay to an unaddressed state, i.e., ``photon budget.'' To slow SrOH from the molecular beam's peak velocity around 110~m/s to near 0~m/s, $\sim$20,000 photons are scattered. We estimate that an additional $\sim$10,000 photons are scattered in the MOT to trap the molecules. Thus, increasing the photon budget from \tenlasersbudget(\tenlasersbudgetunc) to \fullbudget(\fullbudgetunc) decreases the number of molecules lost to dark states by about an e-folding. Next, slow molecules continue to arrive in the MOT region for $\sim$\fullbudgetloadingtau~ms after the end of laser slowing. The longer MOT lifetime allows molecules loaded early to survive until late-arriving molecules load, thus increasing the number loaded by $\sim12\%$. Finally, we re-optimize the slowing parameters for the improved optical cycle. These factors all contribute to the 4.5$\times$ increase in trapped molecule number due to the added repumpers.

To determine the trapped molecule number, the photon budget itself is calibrated via separate measurements, rather than relying on dispersed laser-induced fluorescence measurements of branching fractions~\cite{LasnerVBRs2022}, as discussed in Appendix~\ref{App:number}. The improvement in the number of trapped molecules without the addition of new repumpers, $N =$\MOTbaseline(\MOTbaselineUnc), compared to the previously reported value of $N=2000(600)$~\cite{LasnerMOT}, is largely due to the increase in the power on several repumping paths, as described in Appendix~\ref{SI:cyclingscheme}. In addition, with the improved calibration of the photon budget obtained in this work, we infer an updated (marginally increased) value of the previously reported number to be $N=2600(500)$~molecules.

Figure~\ref{fig:lifetimes} shows the MOT lifetime increase as the $\X(120)$ repumpers are added to the optical cycle with approximately equal loss rates into $\ell=0$ and 2. For those measurements, the main transition MOT light power was held at \oneVcal~mW. At the lowest MOT light power of \lowPowerHold~mW, we observed a lifetime of 209.2(17.7)~ms, which is limited by loss to unaddressed states (see Figure~\ref{fig:lifetimes_powers}).

\section{Conclusion}

We perform spectroscopy of weak vibronic transitions in a MOT of SrOH by developing repump and depletion spectroscopy methods. We find several states in the $\tilde{A}(020)$ manifold that we then use to identify and repump $\X(120)$ states. Adding these repumpers to the optical cycling scheme, we decrease the loss rate from the optical cycle and increase the number of trapped molecules to $N=\MOTnum(\MOTnumUnc)$. The increase in the number of molecules in the MOT is beneficial for future precision measurements using molecules in an optical dipole trap (ODT).

The use of trapped molecules in this work provides long interaction times, enabling a strong signal for scattering rates on the order of 100~Hz. Furthermore, excited state spectroscopy is facilitated by continuous photon scattering at a rate of $\sim$1~MHz in the MOT, enabling efficient searches for vibrational excited states while maintaining high resolution. The continual scattering also ensures that the molecules sample different vibrational states in this system, allowing the ground state of a transition to be easily assigned across $\sim$10 vibrational manifolds.

In addition, for the first time, we locate the vibrational manifolds that are the key to proposed measurements of UDM with SrOH~\cite{kozyryev2021enhanced} and demonstrate a pathway to directly populate the relevant rovibrational states. Our high-resolution measurement of the $\X(200)$--$\X(03^10)$ spacing confirms the viability of driving the proposed vibrational transitions with microwaves for this precision measurement of temporal variations of $\mu$~\cite{kozyryev2021enhanced}. 

In closely related work, using the results obtained here, SrOH is loaded from a cryogenic buffer gas beam (CBGB) into a red MOT, sub-doppler cooled, and loaded into an ODT. Loading efficiency into the ODT is increased by compression using a conveyor-belt MOT~\cite{Hallas2024, Li2025} to better spatially mode-match the MOT cloud size to the ODT size~\cite{Sawaoka2025}. With the current initial red MOT numbers of $N\sim10^4$, we achieve a trapped molecule number in the ODT of 
$\sim10^3$~\cite{Sawaoka2025}. 
In future experiments, transverse cooling of the molecular beam during slowing can be used to further increase the number of molecules in the red MOT by improving CBGB slowing efficiency~\cite{Alauze2021,Kozyryev2017}. 
This will result in further improvements in the sensitivity to ultralight dark matter using SrOH~\cite{kozyryev2021enhanced}. 

\begin{acknowledgements}
The authors thank Arian Jadbabaie for valuable discussions on molecular structure. This work was done at the Center for Ultracold Atoms (an NSF Physics Frontier Center) and supported by Q-SEnSE: Quantum Systems through Entangled Science and Engineering (NSF QLCI Award OMA-2016244), the Alfred P. Sloan Foundation (G-2023-21036), the Gordon and Betty Moore
Foundation (7947), AOARD: Asian Office of Aerospace Research and Development (FA2386-24-1-4070), and AFOSR: Air Force Office of Scientific Research (DURIP FA9550-24-1-0060).
\end{acknowledgements}

\begin{appendix}

\section{Laser scheme}\label{SI:cyclingscheme}

The optical cycling scheme to scatter on average \fullbudget(\fullbudgetunc) photons before decay to an unaddressed state, i.e. the photon budget, is shown below in Figure~\ref{fig:cycling_diagram}. The overall scheme remains similar to the one originally reported in Ref.~\cite{LasnerMOT}, but contains some changes to optimize the overall scattering rate of the optical cycle. This optimization of the scattering rate allows for a higher number of trapped molecules to a baseline of \MOTbaseline (\MOTbaselineUnc) without the addition of the $\X(120)$ repumpers. Each repumper must decrease at least one vibrational quantum number in order to transfer molecules back into the cycle. Most transitions will have $v_{1,e}-v_{1,g}=\Delta v_1 = -1$ or, if necessary, $\Delta v_2=-1$ or $-2$. In the case of $\X(02^20)$, we use a transition with $\Delta \ell = -2$. 

\begin{figure}[h] 
    \centering
    \includegraphics[width=\columnwidth]{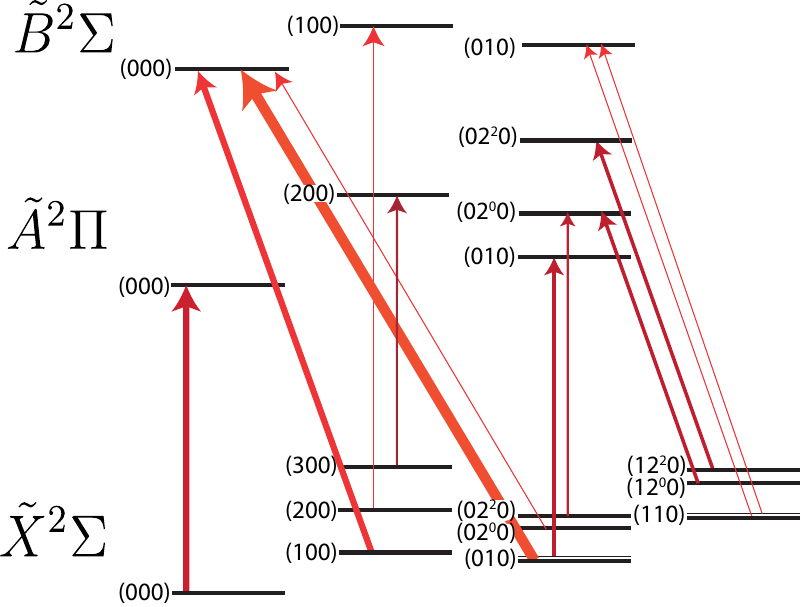} 
    \caption{Full optical cycling scheme. Line thicknesses represent relative laser powers. $\X(010)$ and $\X(110)$ states require separate lasers to address both $N$=1 and 2 states, represented by thicker and thinner lines, respectively. Energies not to scale.}
    \label{fig:cycling_diagram}
\end{figure}

The transitions used in this updated cycling scheme are given in greater detail in Table~\ref{tab:lasers}. The majority of repumpers are around 630~nm (typical of $\Delta v_1=-1$ transitions in the $\X$--$\B$ band), suitable for summed frequency generation (SFG) from Yb-doped ($\sim$1060~nm) and Er-doped ($\sim$1550~nm) fiber amplifiers. These systems also are high power, which is necessary to tolerate the power losses due to combining the different frequencies of light for slowing and trapping. Additionally, as discussed in Ref.~\cite{LasnerMOT}, in order to reach higher scattering rates for systems with many lasers, each repumper must drive molecules out of the ground state much faster (ideally $\gtrsim10\times$ faster) than the molecule falls in. To achieve such scattering rates, higher powers in the repumping lasers are required.

Improvements in the trapped molecule number compared to Ref.~\cite{LasnerMOT} are primarily due to improved laser powers, particularly due to increasing the power of the $\X(010;N=1)$ repumper from 500~mW to 3~W. Additional increases of other repump powers, including $\X(02^00)$ and $\X(200)$, shift the optimal slowing times from 26-30~ms to 24-28~ms, implying a higher effective photon scattering rate. Marginal optimizations to many other experimental parameters also partially contribute to the improved number of trapped molecules.

\begin{table}[]
\resizebox{\columnwidth}{!}{%
\begin{tabular}{|c|c|c|c|c|c|}
    \hline
    \textbf{Ground state ($X^2\Sigma$)} & \textbf{Excited state} & \textbf{Wavelength (nm)} & \textbf{Technology} & \textbf{Power} \\ 
    \hline
    
    $(000)$  & $\A(000)$ & 687.6 & SFG & 800 mW  \\
    $(100)$  & $\B(000)$ & 630.9 & SFG & 1 W  \\
    $(200)$  & $\B(100)$ & 630.5 & SFG* & 100 mW  \\
    $(010; N=1)$  & $\B(000)$ & 624.5 & SFG &  3 W  \\
    $(02^00)$  & $\B(000)$ & 638.0 & SFG & 600 mW  \\
    $(02^20)$  & $\tilde{A}(020)\mu^2\Pi_{1/2}$ & 687.7 & VECSEL &  40 mW  \\
    $(010; N=2)$  & $\tilde{A}(010)\kappa^2\Sigma_{1/2}$ & 674.5 & ECDL & 5 mW  \\
    $(300)$  & $\A(200)$ & 711.5 & ECDL & 6 mW  \\
    $(110; N=1)$ & $\B(010)$ & 629.1 & SFG* & 20 mW  \\
    $(12^00)$ & $\tilde{A}(020)\mu^2\Pi_{1/2}$ & 711.4 & ECDL & 5 mW  \\
    $(12^20)$ & $\tilde{A}(020)\kappa^2\Pi_{1/2}$ & 697.7 & ECDL &  5 mW  \\
    $(110; N=2)$ & $\B(010)$ & 629.2 & SFG* & 5 mW  \\
    \hline
\end{tabular}
}
\caption{Optical cycling scheme achieving \fullbudget(\fullbudgetunc) photons before decay to an unaddressed state. Most lasers are generated by lower-power ECDLs or Summed Frequency Generation (SFG) systems. Daisy-chained SFG systems are noted with an asterisk and described in detail in Ref.~\cite{LasnerMOT}. Powers are measured after all frequencies are combined.}
\label{tab:lasers}

\end{table}

\section{Additional observed vibronic states}\label{SI:otherstates}

All observed transitions are noted in Table~\ref{tab:alltransitions}. Since spectroscopy is not performed up to high $J$ levels in the observed vibronic manifolds, we do not attempt to infer rotational Hamiltonians or determine precise molecular constants. For any excited state used in repumping, it is necessary to determine the rotational state to ensure rotational closure. To estimate the rotational spacings in $\tilde{A}(020)$ states, we use the deperturbed Hamiltonian from Ref.~\cite{CaOHA020} with molecular constants from Ref.~\cite{SrOHA010}, which investigated the SrOH $\tilde{A}(010)$ state. These resulting estimates are consistent with the observed spacings, implying that the Hamiltonian captures the low $J$ spacings in SrOH and the rotational constants in the $\tilde{A}(020)$ state are similar to those in $\tilde{A}(010)$. 

\begin{table}[]
\resizebox{.75\columnwidth}{!}{%
\begin{tabular}{|c|c|}
\hline
State Labels & Energies (cm$^{-1}$) \\ \hline
$\X(03^10; N=1)$ & 1051.350(4) \\ \hline
$\X(12^00; N=1)$ & 1217.875(1) \\ \hline
$\X(12^20; N=2)$ & 1247.105(1) \\ \hline
$\tilde{A}(020; J=1/2)\mu^2\Pi_{1/2}$ & 15275.842(1) \\ \hline
$\tilde{A}(020;J=3/2)\mu^2\Pi_{3/2}$ & 15304.801(1) \\ \hline
$\tilde{A}(020;J=3/2)\kappa^2\Pi_{3/2}$ & 15551.081(1) \\ \hline
$\tilde{A}(020;J=1/2)\kappa^2\Pi_{1/2}$ & 15581.396(1) \\ \hline
\end{tabular}%
}
\caption{Energies of important states for repumping or future precision measurements, determined from transitions in Table~\ref{tab:alltransitions}. Combined with the rotational analysis of $\X(03^10)$ in Ref.~\cite{Fletcher95}, we determine its vibrational energy in the absence of the rotational or $\ell$-doubling contributions to be $T_{03^10}=1051.107(4)~\wn$.}
\label{tab:energies}
\end{table}

We scan for the $\X(000)$--$\tilde{A}(020)$ transitions with around 40~mW of 640~nm light directly overlapping the MOT to drive MOT depletion. When a new excited state is found, it pumps molecules out of the optical cycle, causing depletion of the trapped numbers and lowering the MOT fluorescence. During this process, other transitions in the $\X(v_1,v_2,0) - \tilde{A}^2\Pi_{3/2}(v_1,v_2+2,0)$ band are observed with the same signature of depletion of the MOT fluorescence. This band is additionally overlapped with the $\X(v_1,v_2,0) - \A(v_1+1, v_2,0)$ band in SrOH, allowing for some ambiguity in state assignment, which we resolve as described in Sec.~\ref{AppSec:DLIF}. For each transition, the ground state is determined as described in Sec.~\ref{sec:methods}.

To calibrate the offset in the wavemeter we use to measure laser frequencies, we observe the strontium intercombination line ($^1S_0$--$^3P_1$) in our apparatus with a precision of 0.3~MHz, limited by the finite doppler broadening. By referencing this measurement to the value of 434.829 121 311 (10) THz reported in Ref.~\cite{Ferrari2003}, we determine the wavemeter offset to be +68.8~MHz and report the calibrated frequency in Table~\ref{tab:alltransitions}. Due to drifts in the calibration of a frequency reference during data collection, we estimate the minimum frequency uncertainty in Table~\ref{tab:alltransitions} to be 20~MHz.

\begin{table*}[]
\resizebox{.65\textwidth}{!}{%
\begin{tabular}{|c|c|c|}
\hline
\textbf{Transition frequency (THz)}& \textbf{Ground state} & \textbf{Excited state assignment} \\ \hline
467.103477/587(20)&  $\X(000)$ & $\tilde{A}(020; J=1/2)\kappa^2\Pi_{1/2}$\\ \hline
467.125827/937(20)&  $\X(000)$ & $\tilde{A}(020; J=3/2)\kappa^2\Pi_{1/2}$\\ \hline
467.164325(20)&  $\X(000)$ & $\tilde{A}(020; J=5/2)\kappa^2\Pi_{1/2}$\\ \hline
421.433411(25)&  $\X(12^00)$ & $\tilde{A}(020; J=1/2)\mu^2\Pi_{1/2}$\\ \hline
421.449800(1200)&  $\X(12^00)$ & $\tilde{A}(020; J=3/2)\mu^2\Pi_{1/2}$\\ \hline
429.716296/549(20)&  $\X(12^20)$ & $\tilde{A}(020; J=1/2)\kappa^2\Pi_{1/2}$\\ \hline
429.738520(320)&  $\X(12^20)$ & $\tilde{A}(020; J=3/2)\kappa^2\Pi_{1/2}$\\ \hline
429.698645(100)&  $\X(12^00)$ & $\tilde{A}(020; J=1/2)\kappa^2\Pi_{3/2}$\\ \hline
436.820271(30)&  $\X(02^20)$ & $\tilde{A}(020; J=3/2)\mu^2\Pi_{3/2}$\\ \hline
436.867871(100)&  $\X(02^20)$ & $\tilde{A}(020; J=5/2)\mu^2\Pi_{3/2}$\\ \hline
468.114038(20)&  $\X(110)$ & $\tilde{A}^2\Pi_{3/2}(130)\kappa^2\Delta_{3/2}$,*$^{,\dagger}$ unknown $J$\\ \hline
468.128320(20)&  $\X(010; N=2)$ & $\tilde{A}(030)\kappa^2\Delta_{3/2}$,* unknown $J$\\ \hline
468.120317/422(20)&  $\X(010)$ & $\tilde{A}(030; J=3/2)\kappa^2\Delta_{3/2}$\\ \hline
468.134076/186(20)&  $\X(010)$ & $\tilde{A}(030)\kappa^2\Delta_{3/2}$,*$^{,\dagger}$ unknown $J$\\ \hline
477.658754(55)&  $\X(02^00)$ & $\tilde{A}(220)\kappa^2\Pi_{3/2}$,* unknown $J$\\ \hline
477.675991(20)&  $\X(02^00)$ & $\tilde{A}(220)\kappa^2\Pi_{3/2}$,* unknown $J$\\ \hline
477.614434(26)&  $\X(12^00)$ & $\B(02^00)$, unknown $N$\\ \hline
423.900543(26)&  $\X(03^10)$& $\tilde{A}(010;J=1/2)\kappa^2\Sigma$\\ \hline
423.919453(20)&  $\X(03^10)$& $\tilde{A}(010;J=3/2)\kappa^2\Sigma$\\ \hline
426.409689(20)&  $\X(03^10;N=2)$& $\tilde{A}(020;J=1/2)\mu^2\Pi_{1/2}$\\ \hline
426.870765(20)&  $\X(03^10;N=2)$& $\tilde{A}(010;J=1/2)\kappa^2\Sigma$\\ \hline
423.889685(20)&  $\X(03^10;N=2)$& $\tilde{A}(010;J=3/2)\kappa^2\Sigma$\\ \hline
466.434125/221(20)&  $\X(01^10)$& $\tilde{A}(03^10)\kappa^2\Sigma$, unknown $J$\\ \hline
\end{tabular}

}
\caption{Observed transitions. The rotational state is noted for the excited state if known. The rotational state for each ground state is the lowest odd-parity state unless otherwise noted. The minimum uncertainty noted is 20~MHz due to a conservative estimated drift in the frequency reference used in this work. Asterisks (*) denote states that are assigned using DLIF spectroscopy, described in Appendix Section~\ref{AppSec:DLIF}. For all DLIF measurements, the inferred energy of the ground state populated upon decay is expected to match a known or predicted vibrational state in the $\X$ manifold to within $\sim$3~$\wn$. A dagger ($^\dagger$) denotes states for which the inferred energy of the populated state matches a vibrational state only to within 10~$\wn$, suggesting a small unidentified systematic error. States labeled with daggers are therefore provisional.}
\label{tab:alltransitions}
\end{table*}




\subsection{DLIF spectroscopy to identify excited vibronic states}\label{AppSec:DLIF}

Ambiguity in observed excited-state vibronic labels is removed using dispersed laser-induced fluorescence (DLIF) spectroscopy. DLIF measurements can also reduce the uncertainty of a previously unobserved vibronic state in the $\X$ manifold before attempting high-resolution measurements. Before high-resolution MOT repumping spectroscopy, DLIF spectroscopy is a useful tool to initially narrow the $\X(03^10)$ energy so that MOT spectroscopy can proceed more quickly. 

The apparatus and methodology of our DLIF are as follows. Cold ($\sim$4~K) SrOH is created in a cryogenic buffer gas beam source, and then spectroscopy is performed 1~inch from the cell aperture using a retro-reflected probe laser traveling perpendicular to the molecular beam axis, as described in Ref.~\cite{LasnerVBRs2022}. The fluorescence from the molecules is collected by a lens and curved mirror, after which the photons are dispersed onto an EMCCD by a Czerny-Turner-style spectrometer.

\begin{figure} 
    \centering
    \includegraphics[width=.8\columnwidth]{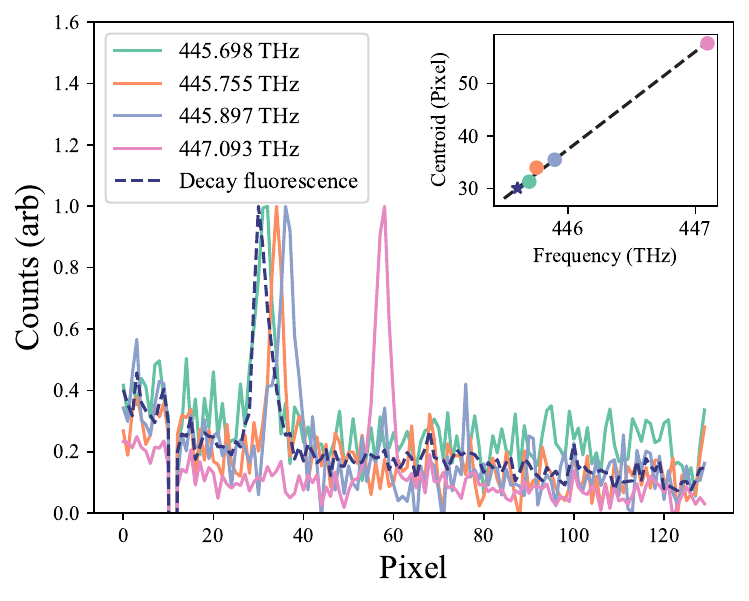} 
    \caption{Fluorescence from a decay in SrOH compared to narrow-line light from a laser. Inset shows the fit centers in pixels versus the known frequency. The line is fit to the known frequencies and used to calibrate the decay fluorescence.}
    \label{fig:DLIF}
\end{figure}

The fluorescence observed is from molecules driven from the natural population in the $\X$ manifold, without any state preparation. We drive the transition of interest, using the same light source from MOT spectroscopy, and observe the decay fluorescence frequency. The horizontal pixel scale is calibrated to frequency by dispersing narrow-band ($\sim$2~MHz) laser light at several different, known frequencies, determining the center of the resulting peak, and performing a linear fit, as in Figure~\ref{fig:DLIF}. The peak centers are fit using a center-of-mass fit procedure. This line is used to determine the decay fluorescence frequency, after fitting the peak center, to a highest possible resolution of $\sim$0.5~px, which corresponds to $\sim$1$\wn$ uncertainty.
From spectroscopy in the MOT, the ground-state assignment $\X(v_1^g,v_2^g,\ell^g,v_3^g)$, labeled $x_g'$, and transition energy are already known. The decay fluorescence is then measured on the spectrometer. For an excited electronic manifold $M=\tilde{A}^2\Pi$ or $\B$ and vibrational quantum numbers $(v_1^e,v_2^e,\ell^e,v_3^e)$, where $M(v_1^e,v_2^e,\ell^e,v_3^e)$ is labeled $x_e$, the strongest fluorescence, typically $\sim90\%$ of the total, accompanies the decay to $\X(v_1^e,v_2^e,\ell^e,v_3^e)$, labeled $x_g$. Thus, assigning the ground state from the dominant decay fluorescence assigns the excited-state vibronic quantum numbers.

To assign $x_g$ and thus $x_e$, the energy of the probe light is added to the energy of the assigned ground state to give the energy of the excited state: $E_{x_e}=E_{x_g'-x_e}+E_{x_g'}$. Next, after the decay fluorescence energy $(E_{x_e - x_g})$ is measured, then the ground state energy can be calculated as $E_{x_g}=E_{x_e}-E_{x_e-x_g}$. The energy $E_{x_g}$ is sufficient to determine its vibronic label $\X(v_1^e,v_2^e,\ell^e,v_3^e)$, which then assigns $x_e$ as well.

In the case of $\tilde{A}(030;J=3/2)\kappa^2\Delta_{3/2}$, which is predominately of $\ell=3$ character, the rotational state is also assigned. First, the vibronic assignment is confirmed in the usual way from the $\X(01^10;N=1)$--$\tilde{A}(030)\kappa^2\Delta_{3/2}$ transition, as described in Section~\ref{sec:methods}. Then, the rotational state is further assigned by driving $\X(01^10; N=2)$--$\tilde{A}(030)\kappa^2\Delta_{3/2}$ and observing the parity doublet splitting of $\sim71$~MHz on top of the spin-rotation splitting of 181~MHz in the ground state. Seeing transitions from both $J=1/2$ and 3/2 split by the known spin-rotation splitting confirms that the excited state must be either $J=1/2$ or 3/2, due to the selection rule of $\Delta J=0,1$. Observation of the parity doublet splitting implies that the excited state parity is unresolved, since the parity doublets are resolved in the ground states. In addition, the decay fluorescence is consistent with the energy of $\X(03^30)$~\cite{Fletcher95}. Together, the known spacings and selection rules confirm the assignment as $\tilde{A}(030)\kappa^2\Delta_{3/2}$.
This observed excited state is approximately $(3^2-1^2)g_{22}=8*7.55\wn$ above the $\tilde{A}(030)\kappa^2\Sigma_{1/2}$ state, of mostly $\ell=1$ character, up to rotational energies and perturbations to these states by Renner-Teller interactions. Thus, assigning the $\X(01^10;N=1)$--$\tilde{A}(030)\kappa^2\Delta_{3/2}$ state provided a helpful starting point for $\X(01^10;N=1)$--$\tilde{A}(030)\kappa^2\Sigma_{1/2}$ transition frequency estimates. 

Additionally, the spectrometer is used to provide an initial, lower-resolution energy for $\X(03^10).$ For this measurement, we drive the transition $\X(01^10;N=1)$--$\tilde{A}(010;J=1/2)\kappa^2\Sigma_{1/2}$ and average the dispersed fluorescence on the spectrometer to see the $\Delta v_2=+2$ decay to $\X(03^10).$ The branching fractions are expected to be similar to the observed $\A(000)$--$\X(020)$ decays of $\sim$0.03\%~\cite{LasnerVBRs2022}. The decay is observed and confirmed to be $\sim2\omega_2\approx 2*377\wn$ redder than the main diagonal decay. We expect the rovibronic decay to primarily populate $\X(03^10;N=1)$ but contain a non-negligible amplitude to $\X(03^10;N=2)$. The observed decay that is 2(1)~$\wn$ above $\X(200;N=1)$ is consistent with the energy determined at high resolution of $\X(03^10;N=1)$, which lies 2.275(4)$\wn$ above $\X(200;N=1)$.

In two cases, noted in Table~\ref{tab:alltransitions}, the measured energy $E_{x_g}$ did not agree with any known or predicted energies for excited vibrational levels in the $\X$ manifold within the expected uncertainty. The vibrational state $x_g$ is still significantly closer to one known/predicted state than any other (i.e. 10$\wn$ from the assigned state but >50$\wn$ from any other). Regardless, to account for the higher error, an additional asterisk is given. 

\subsection{Repumping $\X(12^00)$ through $B(02^00)$}

Initially, one potential pathway for repumping $\X(12^00)$ involved driving it through the $B$ state with the $\X(12^00)$--$\B(02^00)$ transition, since the excited state had already been observed in Ref.~\cite{SrOH020}. As we scan the frequency of a narrow-band dye laser around the $\X(12^00)$--$\B(02^00)$ band at 627~nm, resonance should appear as increased MOT fluorescence. Some repumping is found at \XBrepumpcm, but the increase in MOT fluorescence is $\sim10\times$ weaker than the increase in fluorescence from the $\X(12^00) - \tilde{A}(020)\mu^2\Pi_{1/2}$ repumping pathway. Furthermore, only one rotational state is found, despite broad scans around the observed resonance. 

It remains unclear why this repumping pathway is less successful than repumping $\X(12^00)$ through $\tilde{A}(020)\mu^2\Pi_{1/2}$. To study the $\B(02^00)$ state further using the MOT, $\X(02^00)$--$\B(02^00)$ can be driven for depletion spectroscopy, which would provide higher SNR for detecting rovibronic states in the $\B(02^00)$ manifold, compared to repumping measurements. The measured energy of $\B(02^00)$, subtracting the rotational energy using constants reported in Ref.~\cite{SrOH020}, is approximately $17151.8$~cm$^{-1}$, which is in some disagreement with the value of \Bvalpaper~cm$^{-1}$ from Ref.~\cite{SrOH020}. The difference  cannot be resolved by assuming the excited state is in fact $\B(02^20)$, and the source of the disagreement is unclear. 
Additionally, despite $\B(200)$ being predicted, significant spectroscopy efforts had previously failed to find it as a repumping pathway for $\X(300)$, as discussed in Ref.~\cite{LasnerMOT}. Due to the inability to use either $\B(02^00)$ or $\B(200)$ as states for effective repumping, we speculate that there could be some amount of dissociation in high-lying vibrational states within $\B$, which cannot be observed through fluorescence spectroscopy. Depletion-based spectroscopy in the MOT could determine the location of these lines and may be pursued at a later date. 

\section{Spectroscopic line shapes in the MOT}\label{App:lineshape}

To provide insight into the advantages offered by the MOT for broadband spectroscopic searches, we present a simple model of the expected depletion and repumping line shapes. We consider the states involved in a target transition as a lossy two-level system, with $|1\rangle=|g\rangle, |2\rangle=|e\rangle,$ and loss rates $\gamma_1=0, \gamma_2\equiv\gamma.$ We assume that all population begins in the ground state. The resonance frequency is defined to be $\omega_0$, and the transition is driven at frequency $\omega$ such that the detuning is $\Delta=\omega_0-\omega.$ We define $\Omega$ to be the Rabi frequency. This system has analytic solutions~\cite{Shore1990},
\begin{align}
    P_1(t)=&|A|^2 \text{Exp}[-(\gamma-Y)t]+|1-A|^2\text{Exp}[-(\gamma+Y)t]  \label{eq:P1} \\
    &+2|A(1-A)|\cos(\Omega't+\phi_1)\text{Exp}[-\gamma t] \notag \\
    P_2(t)=&|B|^2\text{Exp}[-(\gamma-Y)t]+|B|^2\text{Exp}[-(\gamma+Y)t] \label{eq:P2} \\
    &-2|B|^2\cos(\Omega't)\text{Exp}[-\gamma t] \notag,
\end{align}
where
\begin{align*}
    A&= \frac{\Omega'-\Delta+i(Y+\gamma)}{2(\Omega'+iY)},\\
    B&= \frac{\Omega}{2(\Omega'+iY)},\\
    Y&= 1/\sqrt{2} [(X^2+4\Delta^2\gamma^2)^{1/2}-X]^{1/2},\\
    \Omega'&= 1/\sqrt{2} [(X^2+4\Delta^2\gamma^2)^{1/2}+X]^{1/2},\\
    X &= |\Omega|^2 + \Delta^2 - \gamma^2,
\end{align*}
and $\phi_1$ is the complex phase of $A(1-A)$. 

We are interested in the population after some interaction time $t,$ which is typically on the scale of ms or longer when performing spectroscopy in the MOT. In this case, $\gamma t\gg1,$ where typically $\gamma\approx2\pi\times7$~MHz in the excited electronic states of SrOH. Furthermore, $Y>0$ so terms proportional to both $\exp[-\gamma t]$ and $\exp[-(\gamma+Y)t]$ are vanishingly small. Then Eqs.~\ref{eq:P1}--\ref{eq:P2} become
\begin{align*}
P_1(t) \approx &|A|^2 e^{- (\gamma - Y) t} \\
P_2(t) \approx  & |B|^2 e^{-(\gamma - Y) t}.
\end{align*}

Expanding $Y^2$ in the large detuning limit where $\gamma, \Omega \ll \Delta$ to second order results in $Y^2 \rightarrow \gamma^2(1-\Omega^2/\Delta^2)$. Taking $Y \rightarrow \gamma \sqrt{1-\Omega^2/\Delta^2}\approx \gamma(1-\Omega^2/(2 \Delta^2))$, the populations are
\begin{align*}
    P_1(t) \approx & |A|^2 e^{-\frac{\gamma}{2}\frac{\Omega^2}{\Delta^2}t}\\
    P_2(t) \approx & |B|^2 e^{-\frac{\gamma}{2}\frac{\Omega^2}{\Delta^2}t}.
\end{align*}
The total population moved outside of the two-state system is
$$P_\text{lost}(t) = 1 - (P_1 + P_2) \approx 1 - (|A|^2+|B|^2)e^{-\frac{\gamma}{2}\frac{\Omega^2}{\Delta^2}t}.$$
Finally, note that $P_\text{lost}(0) = 0$, so by inspection $|A|^2+|B|^2=1$ and the result simplifies to Eq.~\ref{eq:numremoved}: $$P_\text{lost}(t) \approx 1-e^{-\frac{\gamma}{2}\frac{\Omega^2}{\Delta^2}}.$$

As discussed in Section~\ref{sec:methods}, this model accounts for the non-zero repumping we observe for detunings 6~GHz from the $\X(12^20) - \tilde{A}(020)\kappa^2\Pi_{1/2}$ resonance.

\section{High-resolution measurement of $\X(03^10)$ energy for a $\mu$-variation probe}\label{App:sciencestates}

Both $\X(200)$ and $\X(03^10)$ were observed previously in Refs~\cite{Presunka95} and \cite{Fletcher95}, but the absolute energy of $\X(03^10)$ had not been measured. In this work, the $N=1$ and 2 states of $\X(03^10)$ are both observed to high resolution and found to be consistent with the rotational spacings reported in Ref.~\cite{Fletcher95}.

During this work using the apparatus described in Appendix Section~\ref{AppSec:DLIF}, the $\X(03^10)$ state is measured initially using DLIF on a $\Delta v_2=+2$ decay from $\X(01^10;N=1) - \tilde{A}(010;J=1/2)\kappa^2\Sigma_{1/2}$ and found to be $2(1)~\wn$ above $\X(200;N=1)$. Although we have not made a definitive assignment of the rotational state observed in $\tilde{A}(030)\kappa^2\Sigma_{1/2}$, it can still be used to selectively populate $\X(03^10),$ which is a vibrational manifold of interest for probing $\mu$-variation.
After depleting the MOT into $\X(03^10),$ the repumping light is tuned to the prediction for $\X(03^10) - \tilde{A}(010)\kappa^2\Sigma_{1/2}$ and scanned. The $\X(03^10; N=1,2)$ levels are observed and repumped through $\tilde{A}(010; J=1/2,3/2)\kappa^2\Sigma_{1/2}$. Furthermore, the transition $\X(03^10; N=2)$--$\tilde{A}(020;J=1/2)\mu^2\Pi_{1/2}$ is driven. 

We report the energy of $\X(03^10)$ by subtracting the rotational energy ($1B$) from the measured energy of $N=1$, using the spectroscopic constants from Ref.~\cite{Fletcher95}. The predicted spin-rotation splitting of \SRsplitting~MHz is not observed. We observe only one parity, and the splitting of the parity doublet is predicted to be $\sim$24~MHz in $N=1$. Thus, although the transition was measured to an uncertainty of <10~MHz, we report an uncertainty of 120~MHz on the energy to account for the lack of $J$ assignment in the ground state.

\begin{figure} 
    \centering
    \includegraphics[width=.8\columnwidth]{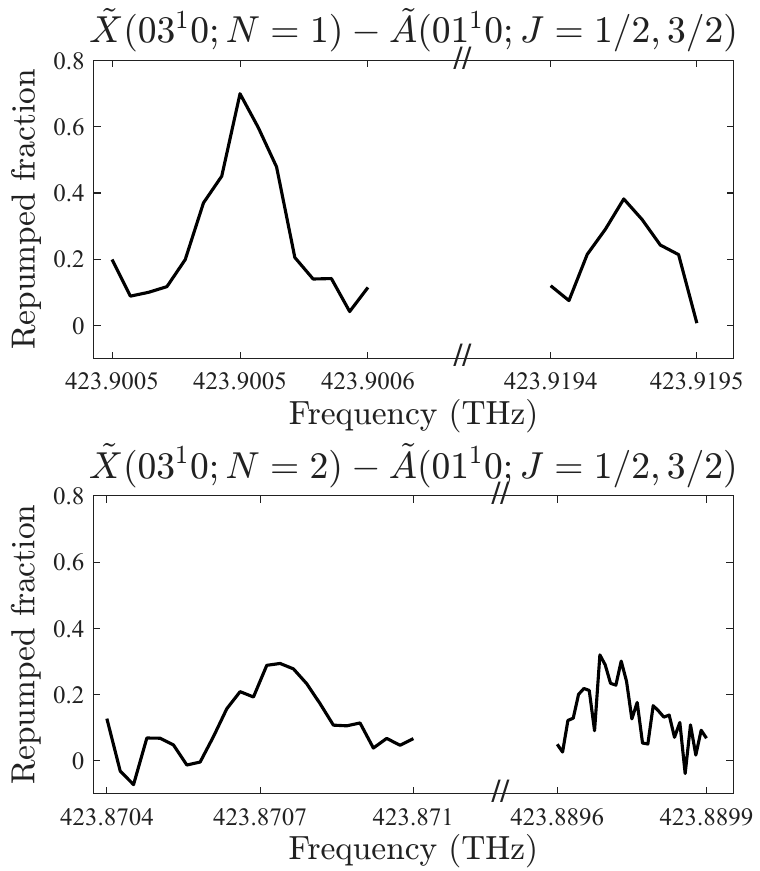} 
    \caption{Repumped fraction of molecules after depleting the MOT into $\X(03^10)$. The upper plot shows $\X(03^10;N=1)$ repumped through the lowest two rotational levels in $\tilde{A}(01^10)$ (here the excited state is written explicitly with its predominate $\ell=1$ character), separated by the known rotational spacing in the $\tilde{A}$ state. The lower plot shows the $\X(03^10;N=2)$ level repumped through the same excited states. These data confirm the assignment of the rotational levels in $\X(03^10)$ and are consistent with Ref.~\cite{Fletcher95}.}
    \label{fig:X0310}
\end{figure}

\begin{figure} 
    \centering
    \includegraphics[width=.8\columnwidth]{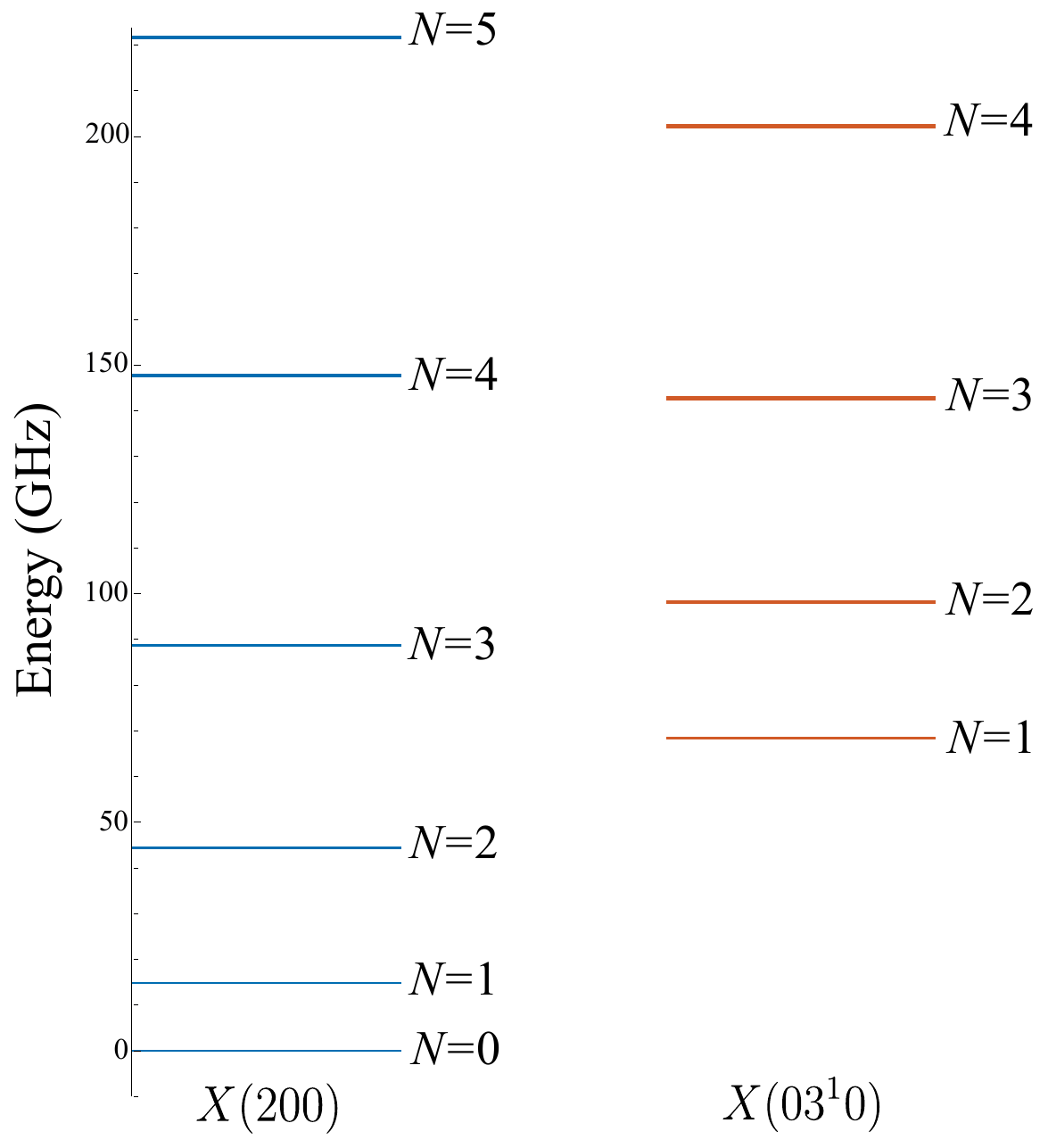} 
    \caption{Diagram of rotational spacings in $\X(200)$ and $\X(03^10)$ using the measured $\X(03^10; N=1, 2)$ levels. The energy of $\X(200; N=0)$ is defined as zero. Spin rotation and $\ell$-doubling are included in the Hamiltonian when appropriate but the spacings are not visible on this scale.}
    \label{fig:rotSpacing}
\end{figure}

With these assignments, the spacings between the low-lying rotational levels in $\X(03^10)$ and $\X(200)$, which are of interest for $\mu$-variation measurements, are calculated. The ladder of low-lying rotational states in Figure~\ref{fig:rotSpacing} illustrates the many options for transitions in the $\X(200)$--$\X(03^10)$ band that lie in the microwave regime. These measurements confirm the estimates in Ref.~\cite{kozyryev2021enhanced}.

\section{Characterization of improved optical cycle}\label{App:number}

With the addition of the $\X(12^00)$ and $\X(12^20)$ repumpers, the average number of photons cycled before $(1-1/e)$ molecules are lost to unaddressed states is improved. We define this as the `photon budget' and label it as $N_b.$ It is important to have an accurate measurement or estimate of the photon budget since it is used to calibrate the trapped molecule number in the MOT. The original DLIF measurements in Ref.~\cite{LasnerVBRs2022} had estimated the size of the $\X(120)$ decays, but the decays were near the noise level of the measurement.
Additionally, these estimations of the photon budget assumed that the measured decays comprised all possible decays. This model holds when only stronger decays are addressed but the model is no longer accurate when the weaker measured decays are also addressed since they are expected to be a similar strength as the unmeasured decays. Thus, adding the $\X(120)$ repumps makes it necessary to determine the photon budget with a new method, which we describe here. 

\subsection{Impact of photon budget on MOT lifetime}

\begin{figure}[h] 
    \centering
    \includegraphics[width=.8\columnwidth]{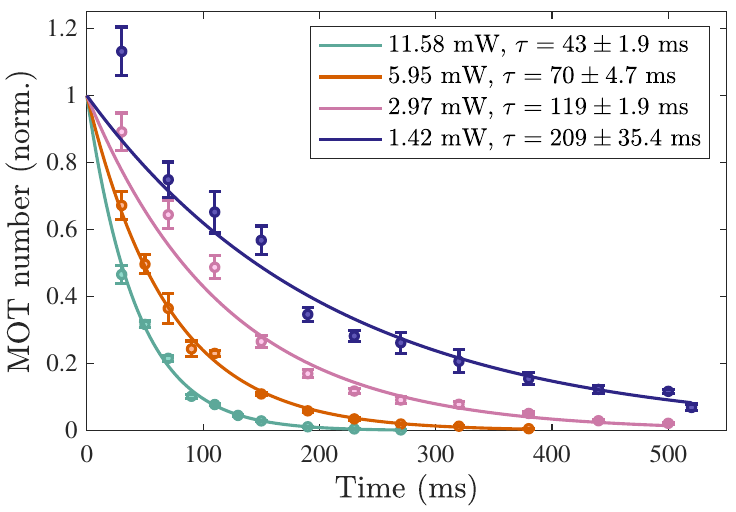} 
    \caption{Extension of the lifetime with lower MOT holding powers, with initial population normalized. At lower hold powers, the molecules scatter more slowly and so remain in the optical cycle for longer. This is consistent with the MOT lifetime being limited by decay to unaddressed states.}
    \label{fig:lifetimes_powers}
\end{figure}

Data on the MOT lifetimes as a function of MOT light beam power and closure of the optical cycle is consistent with the MOT lifetime continuing to be limited by decays to unaddressed states. With all 12 repumpers, the lifetime can be extended to \maxlifetime(\maxlifetimeUnc)~ms, beyond the previous best lifetime with 10 repumpers of \maxlifetimeten(\maxlifetimetenUnc)~ms. MOT lifetimes also scale with the beam powers, as in Figure~\ref{fig:lifetimes_powers}, since the power determines the scattering rate and therefore sets the speed with which molecules fall into unaddressed states.

\subsection{Determination of trapped molecule number}

The MOT fluorescence is directly proportional to the number of trapped molecules, $N_m$. To calculate the number accurately, we use the following process. We start by considering the number of counts on the imaging camera $N_c=S \gamma_\text{det},$ where $S$ is the sensitivity of the camera at the detection wavelength of 611~nm and $\gamma_\text{det}$ is the total number of detected photons.

We also determine $\gamma_\text{det}=\gamma_\text{emit}f_\gamma f_\text{SA},$ where $\gamma_\text{emit}$ is the total number of emitted photons, $f_\gamma$ is the proportion of detectable photons (within the bandwidth of the spectral filters used) and $f_\text{SA}$ is the fraction of photons incident on the optical imaging system, calculated using the NA given by a Zemax simulation of the imaging system. The imaging system includes bandpass filters to detect fluorescence at 611~nm, emitted from $\B(v_1v_2v_3)\rightarrow\X(v_1v_2v_3)$ decays, free from background arising from any laser. From known branching fractions~\cite{LasnerVBRs2022}, we compute $f_{\gamma}\approx$~4\%, primarily due to decay from $\B(000)$ due to the $\X(100)$--$\B(000)$ repumping transition. Finally, the number of emitted photons is $\gamma_\text{emit}=N_m N_b,$ where $N_m$ is finally the number of molecules in the MOT and $N_b$ is the photon budget. Therefore,
\begin{equation}
    N_m=\frac{N_c}{S N_b f_\gamma f_\text{SA}}.
    \label{eq:num}
\end{equation}

This expression assumes that a) there are effectively no untrapped molecules in the image, and b) the MOT is imaged for much longer than the MOT $1/e$  lifetime. A full procedure to calibrate the trapped molecule number, accounting for finite MOT lifetime and temporally extended loading, is discussed in Refs.~\cite{Williams2017, LasnerMOT}. 

The number of molecules trapped in the MOT is measured by imaging the MOT with a delay of 45~ms after the end of slowing, to ensure that any untrapped molecules have passed out of the MOT region. The lifetime of the MOT at this holding power is measured to be \numlifetime(\numlifetimeunc)~ms. The loading time of the MOT is \fullbudgetloadingtau(\fullbudgetloadingtauunc)~ms. These are used to calculate the peak number trapped, as in Ref.~\cite{LasnerMOT}.

\subsection{Determining the photon budget}

As can be seen in Equation~\ref{eq:num}, the overall computed number of trapped molecules, $N_m$, depends inversely on the photon budget, $N_b$. 
In previous work~\cite{LasnerMOT}, uncertainty in $N_m$ was dominated by the estimate of $N_b$, derived from measurements of vibronic branching fractions relevant to the optical cycle~\cite{LasnerVBRs2022}. Here, we describe a direct measurement of $N_b$ as a function of the number of optical cycling lasers. 

In Equation~\ref{eq:num}, $N_c \propto 1/N_b$, provided that the MOT is imaged to $\sim$4 $e$-foldings and no untrapped molecules are imaged. Using this, we can relate the ratio of MOT fluorescence increase to photon budget increase. Since the photon budget is known with high certainty for lower closure (corresponding to lower photon budget and lower number of repumping lasers), we can use the measured brightness increase to calculate the photon budget increase beyond the well-known value.

Specifically, for some number of repumpers, $i$, corresponding to photon budget, $N_{b_i}$, and with measured fluorescence, $N_{c_i}$:
$$N_{c_i} = \frac{N_{m}}{S N_{b_i} f_\gamma f_{SA}}.$$
Then to compare the fluorescence for some number of repumping lasers where $i>3$ to the case with 3 lasers, we gather all constant terms to one side and get:
\begin{equation}
    N_{c_i}N_{b_i} =\frac{N_{m}}{S f_\gamma f_{SA}} = N_{c_3}N_{b_3}.
\end{equation}\label{eq:photonbudgetassumptions}

Thus, the increase in fluorescence can be directly related to the increase in the photon budget, with 
\begin{equation}
    \frac{N_{c_i}}{N_{c_3}}=\frac{N_{b_i}}{N_{b_3}}.
    \label{eq:photonbudgetratio}
\end{equation}

The photon budget is $N_{b_i} = 1/\ell_i = 1/(\tilde{\ell}_i+\ell'),$ where $\tilde{\ell}$ is the loss probability to unaddressed states estimated from DLIF measurements, and $\ell'$ is an additional loss probability due to decays beneath the noise floor of previous measurements. For each set of $i$ lasers, we have calculated $\tilde{\ell}_i$ with a Markov-Chain model from the DLIF data~\cite{LasnerVBRs2022}. Thus, the total loss for a set of $i$ lasers in the cycle is 
\begin{equation}
    \ell_i = \tilde{\ell}_i + \ell'.
    \label{eq:ell}
\end{equation}
Then, substituting Equation~\ref{eq:ell} into Equation~\ref{eq:photonbudgetratio} and solving for $\ell'$ yields
\begin{equation}
    \ell'= \frac{N_{c_i} \tilde{\ell_i} - N_{c_3} \tilde{\ell_3}}{N_{c_3}-N_{c_i}}.
    \label{eq:elltildeFitting}
\end{equation}

\begin{figure}[h] 
    \centering
    \includegraphics[width=.8\columnwidth]{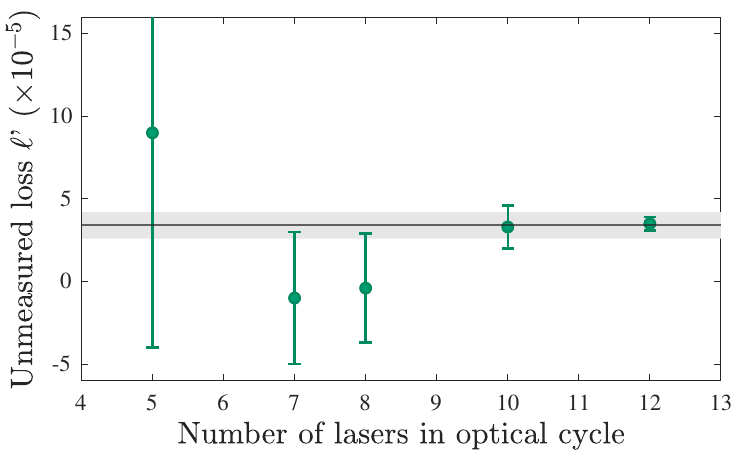} 
    \caption{Values for $\ell'$ from measurements of MOT fluorescence with different numbers of lasers in the optical cycle, relative to $i=3$, calculated with  Equation~\ref{eq:elltildeFitting}. The line shows the weighted average for $\ell'$ of all 5 measurements with the 2$\sigma$ uncertainty represented by the shading. The final value of $\ell'=3.4(8)\times10^{-5}$ is primarily determined by the last two measurements.}
    \label{fig:elltilde}
\end{figure}

MOT fluorescence is measured for several sets of lasers within the total optical cycle as listed in Table~\ref{tab:photonbudget}, with the MOT fluorescence with $i=3$ recorded concurrently as a direct reference for each measurement, since for $i=3$ $\tilde{\ell}_3\gg\ell'$and so $\ell_3\approx\tilde{\ell}_3$. Thus, there are 5 measurements and each yields a value for the additional decay $\ell'$ through Equation~\ref{eq:elltildeFitting}, all given in Figure~\ref{fig:elltilde}. Each $\ell'$ calculation based on the measurements also incorporates the previously measured loss probability through known channels for an optical cycle with $i$ repumpers, $\tilde{\ell_i}$.

A weighted average of all measurements gives $\ell'=3.4(8)\times10^{-5}$. As in Table~\ref{tab:photonbudget}, the estimated loss based on DLIF measurements gives $\tilde{\ell}_{12}=3.33(20)\times10^{-5}$. These values are roughly the same size, implying that the loss channels \textit{not} observed in the original DLIF measurements add up to be significant relative to the smallest observed decays. While the DLIF measurements are a critical and useful tool for designing the initial cycling scheme, the unmeasured loss channels lead to significant error in photon budget estimations, after most measured loss channels are addressed.

\begin{table*}[]
\resizebox{.8\textwidth}{!}{%
\begin{tabular}{|c|c|c|c|c|}
\hline
\textbf{Num. repumpers (i)} & \textbf{Lasers added} & \textbf{$\tilde{\ell_i}$} & $\tilde{\ell_i} + \ell'$ & \textbf{$\gamma=1/\left(\tilde{\ell_i}+\ell'\right)$} \\ \hline\hline
3 & (000), (100), (200) & $1.04(4)\times10^{-3}$ & $1.07(4)\times10^{-3} $ & 933(35) \\ \hline
5 & $\dots,(010; N=1,2)$ & $6.03(36)\times10^{-4}$ & $6.4(4)\times10^{-4}$ & 1522(82) \\ \hline
7 & $\dots, (02^00), (02^20)$ & $2.22(29)\times10^{-4}$ & $2.56(30)\times10^{-4}$ & 4603(226) \\ \hline
8 & $\dots, (300)$ & $1.52(27)\times10^{-4}$ & $1.86(28)\times10^{-4}$ & 6509(313) \\ \hline
10 & $\dots, (110; N=1, 2)$ & $6.84(1.05)\times10^{-5}$ & $1.02(13)\times10^{-4}$ & 9813(461) \\ \hline
12 & $\dots, (12^00), (12^20)$ & $3.33(20)\times10^{-5}$ & $6.7(8)\times10^{-5}$ & 14576(669) \\ \hline
\end{tabular}%
}
\caption{The weighted average from the 5 measurements gave $\ell'=3.4(8)\times10^{-5}$. The uncertainties for the total photon budgets only reflect the \textit{statistical} uncertainties, as discussed in this Appendix, the estimated systematic uncertainties are conservatively on the order of 5\%.}
\label{tab:photonbudget}
\end{table*}

This measurement requires that both conditions have the same number of trapped molecules and that all trapped molecules are imaged. To ensure this assumption holds, we pump all molecules into the lowest two vibrational ground states, $\X(000)$  and $\X(100)$. If all trapped molecules are pumped into those two states, then all molecules can be imaged regardless of the optical cycle. Molecules are pumped into these states by first sub-Doppler cooling to $\lesssim$100~$\mu$K using a $\Lambda$-enhanced gray molasses scheme described in Ref.~\cite{Sawaoka2025}. Then, all repumping lasers, other than those addressing $\X(000)$ and $\X(100)$, are turned on. Finally, the molecules are recaptured in the MOT and the fluorescence with two different optical cycles is measured. We image for 100~ms total. With the full optical cycle of 12 lasers, the trapped molecule lifetime is 24(3.2)~ms. Thus, imaging for 100~ms ensures that the MOT fluorescence is imaged past 4 $e$-foldings. 

The most significant systematic error in this measurement arises from not completely imaging the trapped molecules and comprises $\sim$2\% of the overall number. Additionally, though most decays from the $B$ state are due to repumping $\X(100),$ other ground states are also repumped through $B$ and thus are detectable, adding an additional uncertainty of 0.2\%, to account for the slight differences in $f_{\gamma}$ not incorporated into the model. Ultimately, we conservatively estimate a systematic uncertainty of 5\% and add it in quadrature to the known statistical uncertainty of 5\%, giving a total uncertainty in the photon budget of 7\%.


\end{appendix}

\bibliographystyle{apsrev4-2}
\bibliography{sroh}

\begin{thebibliography}{46}%
\makeatletter
\providecommand \@ifxundefined [1]{%
 \@ifx{#1\undefined}
}%
\providecommand \@ifnum [1]{%
 \ifnum #1\expandafter \@firstoftwo
 \else \expandafter \@secondoftwo
 \fi
}%
\providecommand \@ifx [1]{%
 \ifx #1\expandafter \@firstoftwo
 \else \expandafter \@secondoftwo
 \fi
}%
\providecommand \natexlab [1]{#1}%
\providecommand \enquote  [1]{``#1''}%
\providecommand \bibnamefont  [1]{#1}%
\providecommand \bibfnamefont [1]{#1}%
\providecommand \citenamefont [1]{#1}%
\providecommand \href@noop [0]{\@secondoftwo}%
\providecommand \href [0]{\begingroup \@sanitize@url \@href}%
\providecommand \@href[1]{\@@startlink{#1}\@@href}%
\providecommand \@@href[1]{\endgroup#1\@@endlink}%
\providecommand \@sanitize@url [0]{\catcode `\\12\catcode `\$12\catcode `\&12\catcode `\#12\catcode `\^12\catcode `\_12\catcode `\%12\relax}%
\providecommand \@@startlink[1]{}%
\providecommand \@@endlink[0]{}%
\providecommand \url  [0]{\begingroup\@sanitize@url \@url }%
\providecommand \@url [1]{\endgroup\@href {#1}{\urlprefix }}%
\providecommand \urlprefix  [0]{URL }%
\providecommand \Eprint [0]{\href }%
\providecommand \doibase [0]{https://doi.org/}%
\providecommand \selectlanguage [0]{\@gobble}%
\providecommand \bibinfo  [0]{\@secondoftwo}%
\providecommand \bibfield  [0]{\@secondoftwo}%
\providecommand \translation [1]{[#1]}%
\providecommand \BibitemOpen [0]{}%
\providecommand \bibitemStop [0]{}%
\providecommand \bibitemNoStop [0]{.\EOS\space}%
\providecommand \EOS [0]{\spacefactor3000\relax}%
\providecommand \BibitemShut  [1]{\csname bibitem#1\endcsname}%
\let\auto@bib@innerbib\@empty
\bibitem [{\citenamefont {Kozyryev}\ \emph {et~al.}(2021)\citenamefont {Kozyryev}, \citenamefont {Lasner},\ and\ \citenamefont {Doyle}}]{kozyryev2021enhanced}%
  \BibitemOpen
  \bibfield  {author} {\bibinfo {author} {\bibfnamefont {I.}~\bibnamefont {Kozyryev}}, \bibinfo {author} {\bibfnamefont {Z.}~\bibnamefont {Lasner}},\ and\ \bibinfo {author} {\bibfnamefont {J.~M.}\ \bibnamefont {Doyle}},\ }\href {https://doi.org/10.1103/PhysRevA.103.043313} {\bibfield  {journal} {\bibinfo  {journal} {Phys. Rev. A}\ }\textbf {\bibinfo {volume} {103}},\ \bibinfo {pages} {043313} (\bibinfo {year} {2021})}\BibitemShut {NoStop}%
\bibitem [{\citenamefont {Kozyryev}\ and\ \citenamefont {Hutzler}(2017)}]{kozyryev2017precision}%
  \BibitemOpen
  \bibfield  {author} {\bibinfo {author} {\bibfnamefont {I.}~\bibnamefont {Kozyryev}}\ and\ \bibinfo {author} {\bibfnamefont {N.~R.}\ \bibnamefont {Hutzler}},\ }\href {https://doi.org/https://doi.org/10.1103/PhysRevLett.119.133002} {\bibfield  {journal} {\bibinfo  {journal} {Phys. Rev. Lett.}\ }\textbf {\bibinfo {volume} {119}},\ \bibinfo {pages} {133002} (\bibinfo {year} {2017})}\BibitemShut {NoStop}%
\bibitem [{\citenamefont {Lasner}\ \emph {et~al.}(2025)\citenamefont {Lasner}, \citenamefont {Frenett}, \citenamefont {Sawaoka}, \citenamefont {Anderegg}, \citenamefont {Augenbraun}, \citenamefont {Lampson}, \citenamefont {Li}, \citenamefont {Lunstad}, \citenamefont {Mango}, \citenamefont {Nasir}, \citenamefont {Ono}, \citenamefont {Sakamoto},\ and\ \citenamefont {Doyle}}]{LasnerMOT}%
  \BibitemOpen
  \bibfield  {author} {\bibinfo {author} {\bibfnamefont {Z.~D.}\ \bibnamefont {Lasner}}, \bibinfo {author} {\bibfnamefont {A.}~\bibnamefont {Frenett}}, \bibinfo {author} {\bibfnamefont {H.}~\bibnamefont {Sawaoka}}, \bibinfo {author} {\bibfnamefont {L.}~\bibnamefont {Anderegg}}, \bibinfo {author} {\bibfnamefont {B.}~\bibnamefont {Augenbraun}}, \bibinfo {author} {\bibfnamefont {H.}~\bibnamefont {Lampson}}, \bibinfo {author} {\bibfnamefont {M.}~\bibnamefont {Li}}, \bibinfo {author} {\bibfnamefont {A.}~\bibnamefont {Lunstad}}, \bibinfo {author} {\bibfnamefont {J.}~\bibnamefont {Mango}}, \bibinfo {author} {\bibfnamefont {A.}~\bibnamefont {Nasir}}, \bibinfo {author} {\bibfnamefont {T.}~\bibnamefont {Ono}}, \bibinfo {author} {\bibfnamefont {T.}~\bibnamefont {Sakamoto}},\ and\ \bibinfo {author} {\bibfnamefont {J.~M.}\ \bibnamefont {Doyle}},\ }\href {https://doi.org/10.1103/PhysRevLett.134.083401} {\bibfield  {journal} {\bibinfo  {journal} {Phys. Rev. Lett.}\ }\textbf {\bibinfo {volume} {134}},\ \bibinfo {pages}
  {083401} (\bibinfo {year} {2025})}\BibitemShut {NoStop}%
\bibitem [{\citenamefont {Anderegg}\ \emph {et~al.}(2019)\citenamefont {Anderegg}, \citenamefont {Cheuk}, \citenamefont {Bao}, \citenamefont {Burchesky}, \citenamefont {Ketterle}, \citenamefont {Ni},\ and\ \citenamefont {Doyle}}]{Anderegg2019}%
  \BibitemOpen
  \bibfield  {author} {\bibinfo {author} {\bibfnamefont {L.}~\bibnamefont {Anderegg}}, \bibinfo {author} {\bibfnamefont {L.~W.}\ \bibnamefont {Cheuk}}, \bibinfo {author} {\bibfnamefont {Y.}~\bibnamefont {Bao}}, \bibinfo {author} {\bibfnamefont {S.}~\bibnamefont {Burchesky}}, \bibinfo {author} {\bibfnamefont {W.}~\bibnamefont {Ketterle}}, \bibinfo {author} {\bibfnamefont {K.-K.}\ \bibnamefont {Ni}},\ and\ \bibinfo {author} {\bibfnamefont {J.~M.}\ \bibnamefont {Doyle}},\ }\href {https://www.science.org/doi/abs/10.1126/science.aax1265} {\bibfield  {journal} {\bibinfo  {journal} {Science}\ }\textbf {\bibinfo {volume} {365}},\ \bibinfo {pages} {1156} (\bibinfo {year} {2019})}\BibitemShut {NoStop}%
\bibitem [{\citenamefont {Vilas}\ \emph {et~al.}(2024)\citenamefont {Vilas}, \citenamefont {Robichaud}, \citenamefont {Hallas}, \citenamefont {Li}, \citenamefont {Anderegg},\ and\ \citenamefont {Doyle}}]{Vilas2024}%
  \BibitemOpen
  \bibfield  {author} {\bibinfo {author} {\bibfnamefont {N.~B.}\ \bibnamefont {Vilas}}, \bibinfo {author} {\bibfnamefont {P.}~\bibnamefont {Robichaud}}, \bibinfo {author} {\bibfnamefont {C.}~\bibnamefont {Hallas}}, \bibinfo {author} {\bibfnamefont {G.~K.}\ \bibnamefont {Li}}, \bibinfo {author} {\bibfnamefont {L.}~\bibnamefont {Anderegg}},\ and\ \bibinfo {author} {\bibfnamefont {J.~M.}\ \bibnamefont {Doyle}},\ }\href {https://doi.org/10.1038/s41586-024-07199-1} {\bibfield  {journal} {\bibinfo  {journal} {Nature}\ }\textbf {\bibinfo {volume} {628}},\ \bibinfo {pages} {282} (\bibinfo {year} {2024})}\BibitemShut {NoStop}%
\bibitem [{\citenamefont {DeMille}(2002)}]{Demille2002}%
  \BibitemOpen
  \bibfield  {author} {\bibinfo {author} {\bibfnamefont {D.}~\bibnamefont {DeMille}},\ }\href {https://doi.org/10.1103/PhysRevLett.88.067901} {\bibfield  {journal} {\bibinfo  {journal} {Phys. Rev. Lett.}\ }\textbf {\bibinfo {volume} {88}},\ \bibinfo {pages} {067901} (\bibinfo {year} {2002})}\BibitemShut {NoStop}%
\bibitem [{\citenamefont {Ni}\ \emph {et~al.}(2018)\citenamefont {Ni}, \citenamefont {Rosenband},\ and\ \citenamefont {Grimes}}]{Ni2018}%
  \BibitemOpen
  \bibfield  {author} {\bibinfo {author} {\bibfnamefont {K.~K.}\ \bibnamefont {Ni}}, \bibinfo {author} {\bibfnamefont {T.}~\bibnamefont {Rosenband}},\ and\ \bibinfo {author} {\bibfnamefont {D.~D.}\ \bibnamefont {Grimes}},\ }\href {https://doi.org/10.1039/c8sc02355g} {\bibfield  {journal} {\bibinfo  {journal} {Chemical Science}\ }\textbf {\bibinfo {volume} {9}},\ \bibinfo {pages} {6830} (\bibinfo {year} {2018})}\BibitemShut {NoStop}%
\bibitem [{\citenamefont {Sawant}\ \emph {et~al.}(2020)\citenamefont {Sawant}, \citenamefont {Blackmore}, \citenamefont {Gregory}, \citenamefont {Mur-Petit}, \citenamefont {Jaksch}, \citenamefont {Aldegunde}, \citenamefont {Hutson}, \citenamefont {Tarbutt},\ and\ \citenamefont {Cornish}}]{Sawant2020}%
  \BibitemOpen
  \bibfield  {author} {\bibinfo {author} {\bibfnamefont {R.}~\bibnamefont {Sawant}}, \bibinfo {author} {\bibfnamefont {J.~A.}\ \bibnamefont {Blackmore}}, \bibinfo {author} {\bibfnamefont {P.~D.}\ \bibnamefont {Gregory}}, \bibinfo {author} {\bibfnamefont {J.}~\bibnamefont {Mur-Petit}}, \bibinfo {author} {\bibfnamefont {D.}~\bibnamefont {Jaksch}}, \bibinfo {author} {\bibfnamefont {J.}~\bibnamefont {Aldegunde}}, \bibinfo {author} {\bibfnamefont {J.~M.}\ \bibnamefont {Hutson}}, \bibinfo {author} {\bibfnamefont {M.~R.}\ \bibnamefont {Tarbutt}},\ and\ \bibinfo {author} {\bibfnamefont {S.~L.}\ \bibnamefont {Cornish}},\ }\href {http://dx.doi.org/10.1088/1367-2630/ab60f4} {\bibfield  {journal} {\bibinfo  {journal} {New J. Phys.}\ }\textbf {\bibinfo {volume} {22}} (\bibinfo {year} {2020})}\BibitemShut {NoStop}%
\bibitem [{\citenamefont {Micheli}\ \emph {et~al.}(2006)\citenamefont {Micheli}, \citenamefont {Brennen},\ and\ \citenamefont {Zoller}}]{Micheli2006}%
  \BibitemOpen
  \bibfield  {author} {\bibinfo {author} {\bibfnamefont {A.}~\bibnamefont {Micheli}}, \bibinfo {author} {\bibfnamefont {G.~K.}\ \bibnamefont {Brennen}},\ and\ \bibinfo {author} {\bibfnamefont {P.}~\bibnamefont {Zoller}},\ }\href {https://doi.org/10.1038/nphys287} {\bibfield  {journal} {\bibinfo  {journal} {Nature Physics}\ }\textbf {\bibinfo {volume} {2}},\ \bibinfo {pages} {341} (\bibinfo {year} {2006})}\BibitemShut {NoStop}%
\bibitem [{\citenamefont {Gorshkov}\ \emph {et~al.}(2011)\citenamefont {Gorshkov}, \citenamefont {Manmana}, \citenamefont {Chen}, \citenamefont {Demler}, \citenamefont {Lukin},\ and\ \citenamefont {Rey}}]{Gorshkov2011}%
  \BibitemOpen
  \bibfield  {author} {\bibinfo {author} {\bibfnamefont {A.~V.}\ \bibnamefont {Gorshkov}}, \bibinfo {author} {\bibfnamefont {S.~R.}\ \bibnamefont {Manmana}}, \bibinfo {author} {\bibfnamefont {G.}~\bibnamefont {Chen}}, \bibinfo {author} {\bibfnamefont {E.}~\bibnamefont {Demler}}, \bibinfo {author} {\bibfnamefont {M.~D.}\ \bibnamefont {Lukin}},\ and\ \bibinfo {author} {\bibfnamefont {A.~M.}\ \bibnamefont {Rey}},\ }\href {https://doi.org/10.1103/PhysRevA.84.033619} {\bibfield  {journal} {\bibinfo  {journal} {Phys. Rev. A}\ }\textbf {\bibinfo {volume} {84}},\ \bibinfo {pages} {033619} (\bibinfo {year} {2011})}\BibitemShut {NoStop}%
\bibitem [{\citenamefont {Kobayashi}\ \emph {et~al.}(2019)\citenamefont {Kobayashi}, \citenamefont {Ogino},\ and\ \citenamefont {Inouye}}]{Kobayashi2019}%
  \BibitemOpen
  \bibfield  {author} {\bibinfo {author} {\bibfnamefont {J.}~\bibnamefont {Kobayashi}}, \bibinfo {author} {\bibfnamefont {A.}~\bibnamefont {Ogino}},\ and\ \bibinfo {author} {\bibfnamefont {S.}~\bibnamefont {Inouye}},\ }\href {http://dx.doi.org/10.1038/s41467-019-11761-1} {\bibfield  {journal} {\bibinfo  {journal} {Nat. Comm.}\ }\textbf {\bibinfo {volume} {10}} (\bibinfo {year} {2019})}\BibitemShut {NoStop}%
\bibitem [{\citenamefont {Zeng}\ \emph {et~al.}(2024)\citenamefont {Zeng}, \citenamefont {Deng}, \citenamefont {Yang},\ and\ \citenamefont {Yan}}]{Zeng2024}%
  \BibitemOpen
  \bibfield  {author} {\bibinfo {author} {\bibfnamefont {Z.}~\bibnamefont {Zeng}}, \bibinfo {author} {\bibfnamefont {S.}~\bibnamefont {Deng}}, \bibinfo {author} {\bibfnamefont {S.}~\bibnamefont {Yang}},\ and\ \bibinfo {author} {\bibfnamefont {B.}~\bibnamefont {Yan}},\ }\href {10.1103/PhysRevLett.133.143404} {\bibfield  {journal} {\bibinfo  {journal} {Phys. Rev. Lett.}\ }\textbf {\bibinfo {volume} {133}} (\bibinfo {year} {2024})}\BibitemShut {NoStop}%
\bibitem [{\citenamefont {Christakis}\ \emph {et~al.}(2023)\citenamefont {Christakis}, \citenamefont {Rosenberg}, \citenamefont {Raj}, \citenamefont {Chi}, \citenamefont {Morningstar}, \citenamefont {Huse}, \citenamefont {Yan},\ and\ \citenamefont {Bakr}}]{Christakis2023}%
  \BibitemOpen
  \bibfield  {author} {\bibinfo {author} {\bibfnamefont {L.}~\bibnamefont {Christakis}}, \bibinfo {author} {\bibfnamefont {J.~S.}\ \bibnamefont {Rosenberg}}, \bibinfo {author} {\bibfnamefont {R.}~\bibnamefont {Raj}}, \bibinfo {author} {\bibfnamefont {S.}~\bibnamefont {Chi}}, \bibinfo {author} {\bibfnamefont {A.}~\bibnamefont {Morningstar}}, \bibinfo {author} {\bibfnamefont {D.~A.}\ \bibnamefont {Huse}}, \bibinfo {author} {\bibfnamefont {Z.~Z.}\ \bibnamefont {Yan}},\ and\ \bibinfo {author} {\bibfnamefont {W.~S.}\ \bibnamefont {Bakr}},\ }\href {https://doi.org/10.1038/s41586-022-05558-4} {\bibfield  {journal} {\bibinfo  {journal} {Nature}\ }\textbf {\bibinfo {volume} {614}},\ \bibinfo {pages} {64} (\bibinfo {year} {2023})}\BibitemShut {NoStop}%
\bibitem [{\citenamefont {Holland}\ \emph {et~al.}(2025)\citenamefont {Holland}, \citenamefont {Lu}, \citenamefont {Li}, \citenamefont {Welsh},\ and\ \citenamefont {Cheuk}}]{Holland2025}%
  \BibitemOpen
  \bibfield  {author} {\bibinfo {author} {\bibfnamefont {C.~M.}\ \bibnamefont {Holland}}, \bibinfo {author} {\bibfnamefont {Y.}~\bibnamefont {Lu}}, \bibinfo {author} {\bibfnamefont {S.~J.}\ \bibnamefont {Li}}, \bibinfo {author} {\bibfnamefont {C.~L.}\ \bibnamefont {Welsh}},\ and\ \bibinfo {author} {\bibfnamefont {L.~W.}\ \bibnamefont {Cheuk}},\ }\href {http://dx.doi.org/10.1103/8q8p-mx1l} {\bibfield  {journal} {\bibinfo  {journal} {Phys. Rev. X}\ }\textbf {\bibinfo {volume} {15}} (\bibinfo {year} {2025})}\BibitemShut {NoStop}%
\bibitem [{\citenamefont {Alauze}\ \emph {et~al.}(2021)\citenamefont {Alauze}, \citenamefont {Lim}, \citenamefont {Trigatzis}, \citenamefont {Swarbrick}, \citenamefont {Collings}, \citenamefont {Fitch}, \citenamefont {Sauer},\ and\ \citenamefont {Tarbutt}}]{Alauze2021}%
  \BibitemOpen
  \bibfield  {author} {\bibinfo {author} {\bibfnamefont {X.}~\bibnamefont {Alauze}}, \bibinfo {author} {\bibfnamefont {J.}~\bibnamefont {Lim}}, \bibinfo {author} {\bibfnamefont {M.~A.}\ \bibnamefont {Trigatzis}}, \bibinfo {author} {\bibfnamefont {S.}~\bibnamefont {Swarbrick}}, \bibinfo {author} {\bibfnamefont {F.~J.}\ \bibnamefont {Collings}}, \bibinfo {author} {\bibfnamefont {N.~J.}\ \bibnamefont {Fitch}}, \bibinfo {author} {\bibfnamefont {B.~E.}\ \bibnamefont {Sauer}},\ and\ \bibinfo {author} {\bibfnamefont {M.~R.}\ \bibnamefont {Tarbutt}},\ }\href {http://dx.doi.org/10.1088/2058-9565/ac107e} {\bibfield  {journal} {\bibinfo  {journal} {Quantum Sci. Technol.}\ }\textbf {\bibinfo {volume} {6}} (\bibinfo {year} {2021})}\BibitemShut {NoStop}%
\bibitem [{\citenamefont {Picard}\ \emph {et~al.}(2025)\citenamefont {Picard}, \citenamefont {Park}, \citenamefont {Patenotte}, \citenamefont {Gebretsadkan}, \citenamefont {Wellnitz}, \citenamefont {Rey},\ and\ \citenamefont {Ni}}]{Picard2025}%
  \BibitemOpen
  \bibfield  {author} {\bibinfo {author} {\bibfnamefont {L.~R.~B.}\ \bibnamefont {Picard}}, \bibinfo {author} {\bibfnamefont {A.~J.}\ \bibnamefont {Park}}, \bibinfo {author} {\bibfnamefont {G.~E.}\ \bibnamefont {Patenotte}}, \bibinfo {author} {\bibfnamefont {S.}~\bibnamefont {Gebretsadkan}}, \bibinfo {author} {\bibfnamefont {D.}~\bibnamefont {Wellnitz}}, \bibinfo {author} {\bibfnamefont {A.~M.}\ \bibnamefont {Rey}},\ and\ \bibinfo {author} {\bibfnamefont {K.-K.}\ \bibnamefont {Ni}},\ }\href {https://doi.org/10.1038/s41586-024-08177-3} {\bibfield  {journal} {\bibinfo  {journal} {Nature}\ }\textbf {\bibinfo {volume} {637}},\ \bibinfo {pages} {821} (\bibinfo {year} {2025})}\BibitemShut {NoStop}%
\bibitem [{\citenamefont {Miller}\ \emph {et~al.}(2024)\citenamefont {Miller}, \citenamefont {Carroll}, \citenamefont {Lin}, \citenamefont {Hirzler}, \citenamefont {Gao}, \citenamefont {Zhou}, \citenamefont {Lukin},\ and\ \citenamefont {Ye}}]{Miller2024}%
  \BibitemOpen
  \bibfield  {author} {\bibinfo {author} {\bibfnamefont {C.}~\bibnamefont {Miller}}, \bibinfo {author} {\bibfnamefont {A.~N.}\ \bibnamefont {Carroll}}, \bibinfo {author} {\bibfnamefont {J.}~\bibnamefont {Lin}}, \bibinfo {author} {\bibfnamefont {H.}~\bibnamefont {Hirzler}}, \bibinfo {author} {\bibfnamefont {H.}~\bibnamefont {Gao}}, \bibinfo {author} {\bibfnamefont {H.}~\bibnamefont {Zhou}}, \bibinfo {author} {\bibfnamefont {M.~D.}\ \bibnamefont {Lukin}},\ and\ \bibinfo {author} {\bibfnamefont {J.}~\bibnamefont {Ye}},\ }\href {https://doi.org/10.1038/s41586-024-07883-2} {\bibfield  {journal} {\bibinfo  {journal} {Nature}\ }\textbf {\bibinfo {volume} {633}},\ \bibinfo {pages} {332} (\bibinfo {year} {2024})}\BibitemShut {NoStop}%
\bibitem [{\citenamefont {Burau}\ \emph {et~al.}(2023)\citenamefont {Burau}, \citenamefont {Aggarwal}, \citenamefont {Mehling},\ and\ \citenamefont {Ye}}]{Burau2023}%
  \BibitemOpen
  \bibfield  {author} {\bibinfo {author} {\bibfnamefont {J.~J.}\ \bibnamefont {Burau}}, \bibinfo {author} {\bibfnamefont {P.}~\bibnamefont {Aggarwal}}, \bibinfo {author} {\bibfnamefont {K.}~\bibnamefont {Mehling}},\ and\ \bibinfo {author} {\bibfnamefont {J.}~\bibnamefont {Ye}},\ }\href {https://doi.org/10.1103/PhysRevLett.130.193401} {\bibfield  {journal} {\bibinfo  {journal} {Phys. Rev. Lett.}\ }\textbf {\bibinfo {volume} {130}},\ \bibinfo {pages} {193401} (\bibinfo {year} {2023})}\BibitemShut {NoStop}%
\bibitem [{\citenamefont {Gaul}\ \emph {et~al.}(2024)\citenamefont {Gaul}, \citenamefont {Hutzler}, \citenamefont {Yu}, \citenamefont {Jayich}, \citenamefont {Ilia{\v{s}}},\ and\ \citenamefont {Borschevsky}}]{Gaul2024}%
  \BibitemOpen
  \bibfield  {author} {\bibinfo {author} {\bibfnamefont {K.}~\bibnamefont {Gaul}}, \bibinfo {author} {\bibfnamefont {N.~R.}\ \bibnamefont {Hutzler}}, \bibinfo {author} {\bibfnamefont {P.}~\bibnamefont {Yu}}, \bibinfo {author} {\bibfnamefont {A.~M.}\ \bibnamefont {Jayich}}, \bibinfo {author} {\bibfnamefont {M.}~\bibnamefont {Ilia{\v{s}}}},\ and\ \bibinfo {author} {\bibfnamefont {A.}~\bibnamefont {Borschevsky}},\ }\href {https://doi.org/10.1103/PhysRevA.109.042819} {\bibfield  {journal} {\bibinfo  {journal} {Phys. Rev. A}\ }\textbf {\bibinfo {volume} {109}},\ \bibinfo {pages} {042819} (\bibinfo {year} {2024})}\BibitemShut {NoStop}%
\bibitem [{\citenamefont {Roussy}\ \emph {et~al.}(2023)\citenamefont {Roussy}, \citenamefont {Caldwell}, \citenamefont {Wright}, \citenamefont {Cairncross}, \citenamefont {Shagam}, \citenamefont {Ng}, \citenamefont {Schlossberger}, \citenamefont {Park}, \citenamefont {Wang}, \citenamefont {Ye},\ and\ \citenamefont {Cornell}}]{JILAedm}%
  \BibitemOpen
  \bibfield  {author} {\bibinfo {author} {\bibfnamefont {T.~S.}\ \bibnamefont {Roussy}}, \bibinfo {author} {\bibfnamefont {L.}~\bibnamefont {Caldwell}}, \bibinfo {author} {\bibfnamefont {T.}~\bibnamefont {Wright}}, \bibinfo {author} {\bibfnamefont {W.~B.}\ \bibnamefont {Cairncross}}, \bibinfo {author} {\bibfnamefont {Y.}~\bibnamefont {Shagam}}, \bibinfo {author} {\bibfnamefont {K.~B.}\ \bibnamefont {Ng}}, \bibinfo {author} {\bibfnamefont {N.}~\bibnamefont {Schlossberger}}, \bibinfo {author} {\bibfnamefont {S.~Y.}\ \bibnamefont {Park}}, \bibinfo {author} {\bibfnamefont {A.}~\bibnamefont {Wang}}, \bibinfo {author} {\bibfnamefont {J.}~\bibnamefont {Ye}},\ and\ \bibinfo {author} {\bibfnamefont {E.~A.}\ \bibnamefont {Cornell}},\ }\href {https://doi.org/10.1126/science.adg4084} {\bibfield  {journal} {\bibinfo  {journal} {Science}\ }\textbf {\bibinfo {volume} {381}},\ \bibinfo {pages} {46} (\bibinfo {year} {2023})}\BibitemShut {NoStop}%
\bibitem [{\citenamefont {Andreev}\ \emph {et~al.}(2018)\citenamefont {Andreev}, \citenamefont {Ang}, \citenamefont {DeMille}, \citenamefont {Doyle}, \citenamefont {Gabrielse}, \citenamefont {Haefner}, \citenamefont {Hutzler}, \citenamefont {Lasner}, \citenamefont {Meisenhelder}, \citenamefont {O{'}Leary}, \citenamefont {Panda}, \citenamefont {West}, \citenamefont {West},\ and\ \citenamefont {Wu}}]{ACMEII}%
  \BibitemOpen
  \bibfield  {author} {\bibinfo {author} {\bibfnamefont {V.}~\bibnamefont {Andreev}}, \bibinfo {author} {\bibfnamefont {D.~G.}\ \bibnamefont {Ang}}, \bibinfo {author} {\bibfnamefont {D.}~\bibnamefont {DeMille}}, \bibinfo {author} {\bibfnamefont {J.~M.}\ \bibnamefont {Doyle}}, \bibinfo {author} {\bibfnamefont {G.}~\bibnamefont {Gabrielse}}, \bibinfo {author} {\bibfnamefont {J.}~\bibnamefont {Haefner}}, \bibinfo {author} {\bibfnamefont {N.~R.}\ \bibnamefont {Hutzler}}, \bibinfo {author} {\bibfnamefont {Z.}~\bibnamefont {Lasner}}, \bibinfo {author} {\bibfnamefont {C.}~\bibnamefont {Meisenhelder}}, \bibinfo {author} {\bibfnamefont {B.~R.}\ \bibnamefont {O{'}Leary}}, \bibinfo {author} {\bibfnamefont {C.~D.}\ \bibnamefont {Panda}}, \bibinfo {author} {\bibfnamefont {A.~D.}\ \bibnamefont {West}}, \bibinfo {author} {\bibfnamefont {E.~P.}\ \bibnamefont {West}},\ and\ \bibinfo {author} {\bibfnamefont {X.}~\bibnamefont {Wu}},\ }\href {https://doi.org/10.1038/s41586-018-0599-8} {\bibfield  {journal} {\bibinfo  {journal}
  {Nature}\ }\textbf {\bibinfo {volume} {562}},\ \bibinfo {pages} {355} (\bibinfo {year} {2018})}\BibitemShut {NoStop}%
\bibitem [{\citenamefont {Augenbraun}\ \emph {et~al.}(2020)\citenamefont {Augenbraun}, \citenamefont {Lasner}, \citenamefont {Frenett}, \citenamefont {Sawaoka}, \citenamefont {Miller}, \citenamefont {Steimle},\ and\ \citenamefont {Doyle}}]{Augenbraun2020}%
  \BibitemOpen
  \bibfield  {author} {\bibinfo {author} {\bibfnamefont {B.~L.}\ \bibnamefont {Augenbraun}}, \bibinfo {author} {\bibfnamefont {Z.~D.}\ \bibnamefont {Lasner}}, \bibinfo {author} {\bibfnamefont {A.}~\bibnamefont {Frenett}}, \bibinfo {author} {\bibfnamefont {H.}~\bibnamefont {Sawaoka}}, \bibinfo {author} {\bibfnamefont {C.}~\bibnamefont {Miller}}, \bibinfo {author} {\bibfnamefont {T.~C.}\ \bibnamefont {Steimle}},\ and\ \bibinfo {author} {\bibfnamefont {J.~M.}\ \bibnamefont {Doyle}},\ }\href {https://doi.org/10.1088/1367-2630/ab687b} {\bibfield  {journal} {\bibinfo  {journal} {New J. Phys.}\ }\textbf {\bibinfo {volume} {22}},\ \bibinfo {pages} {022003} (\bibinfo {year} {2020})}\BibitemShut {NoStop}%
\bibitem [{\citenamefont {Hao}\ \emph {et~al.}(2020)\citenamefont {Hao}, \citenamefont {Navr{\'a}til}, \citenamefont {Norrgard}, \citenamefont {Ilia{\v{s}}}, \citenamefont {Eliav}, \citenamefont {Timmermans}, \citenamefont {Flambaum},\ and\ \citenamefont {Borschevsky}}]{Hao2020}%
  \BibitemOpen
  \bibfield  {author} {\bibinfo {author} {\bibfnamefont {Y.}~\bibnamefont {Hao}}, \bibinfo {author} {\bibfnamefont {P.}~\bibnamefont {Navr{\'a}til}}, \bibinfo {author} {\bibfnamefont {E.~B.}\ \bibnamefont {Norrgard}}, \bibinfo {author} {\bibfnamefont {M.}~\bibnamefont {Ilia{\v{s}}}}, \bibinfo {author} {\bibfnamefont {E.}~\bibnamefont {Eliav}}, \bibinfo {author} {\bibfnamefont {R.~G.}\ \bibnamefont {Timmermans}}, \bibinfo {author} {\bibfnamefont {V.~V.}\ \bibnamefont {Flambaum}},\ and\ \bibinfo {author} {\bibfnamefont {A.}~\bibnamefont {Borschevsky}},\ }\href {http://dx.doi.org/10.1103/PhysRevA.102.052828} {\bibfield  {journal} {\bibinfo  {journal} {Phys. Rev. A}\ }\textbf {\bibinfo {volume} {102}} (\bibinfo {year} {2020})}\BibitemShut {NoStop}%
\bibitem [{\citenamefont {Isaev}\ \emph {et~al.}(2010)\citenamefont {Isaev}, \citenamefont {Hoekstra},\ and\ \citenamefont {Berger}}]{Isaev2010}%
  \BibitemOpen
  \bibfield  {author} {\bibinfo {author} {\bibfnamefont {T.~A.}\ \bibnamefont {Isaev}}, \bibinfo {author} {\bibfnamefont {S.}~\bibnamefont {Hoekstra}},\ and\ \bibinfo {author} {\bibfnamefont {R.}~\bibnamefont {Berger}},\ }\href {https://doi.org/10.1103/PhysRevA.82.052521} {\bibfield  {journal} {\bibinfo  {journal} {Phys. Rev. A}\ }\textbf {\bibinfo {volume} {82}},\ \bibinfo {pages} {052521} (\bibinfo {year} {2010})}\BibitemShut {NoStop}%
\bibitem [{\citenamefont {Hanneke}\ \emph {et~al.}(2021)\citenamefont {Hanneke}, \citenamefont {Kuzhan},\ and\ \citenamefont {Lunstad}}]{Hanneke2021}%
  \BibitemOpen
  \bibfield  {author} {\bibinfo {author} {\bibfnamefont {D.}~\bibnamefont {Hanneke}}, \bibinfo {author} {\bibfnamefont {B.}~\bibnamefont {Kuzhan}},\ and\ \bibinfo {author} {\bibfnamefont {A.}~\bibnamefont {Lunstad}},\ }\href {https://doi.org/10.1088/2058-9565/abc863} {\bibfield  {journal} {\bibinfo  {journal} {Quantum Sci. Technol.}\ }\textbf {\bibinfo {volume} {6}},\ \bibinfo {pages} {014005} (\bibinfo {year} {2021})}\BibitemShut {NoStop}%
\bibitem [{\citenamefont {Ferreira}(2020)}]{Ferreira2020}%
  \BibitemOpen
  \bibfield  {author} {\bibinfo {author} {\bibfnamefont {E.~G.~M.}\ \bibnamefont {Ferreira}},\ }\href {https://doi.org/10.1007/s00159-021-00135-6} {\bibfield  {journal} {\bibinfo  {journal} {The Astronomy and Astrophysics Review}\ }\textbf {\bibinfo {volume} {29}},\ \bibinfo {pages} {7} (\bibinfo {year} {2020})}\BibitemShut {NoStop}%
\bibitem [{\citenamefont {Prehn}\ \emph {et~al.}(2016)\citenamefont {Prehn}, \citenamefont {Ibr{\"u}gger}, \citenamefont {Gl{\"o}ckner}, \citenamefont {Rempe},\ and\ \citenamefont {Zeppenfeld}}]{Prehn2016}%
  \BibitemOpen
  \bibfield  {author} {\bibinfo {author} {\bibfnamefont {A.}~\bibnamefont {Prehn}}, \bibinfo {author} {\bibfnamefont {M.}~\bibnamefont {Ibr{\"u}gger}}, \bibinfo {author} {\bibfnamefont {R.}~\bibnamefont {Gl{\"o}ckner}}, \bibinfo {author} {\bibfnamefont {G.}~\bibnamefont {Rempe}},\ and\ \bibinfo {author} {\bibfnamefont {M.}~\bibnamefont {Zeppenfeld}},\ }\href {10.1103/PhysRevLett.116.063005} {\bibfield  {journal} {\bibinfo  {journal} {Phys. Rev. Lett.}\ }\textbf {\bibinfo {volume} {116}} (\bibinfo {year} {2016})}\BibitemShut {NoStop}%
\bibitem [{\citenamefont {Liu}\ \emph {et~al.}(2018)\citenamefont {Liu}, \citenamefont {Hood}, \citenamefont {Yu}, \citenamefont {Zhang}, \citenamefont {Hutzler}, \citenamefont {Rosenband},\ and\ \citenamefont {Ni}}]{Liu2018}%
  \BibitemOpen
  \bibfield  {author} {\bibinfo {author} {\bibfnamefont {L.~R.}\ \bibnamefont {Liu}}, \bibinfo {author} {\bibfnamefont {J.~D.}\ \bibnamefont {Hood}}, \bibinfo {author} {\bibfnamefont {Y.}~\bibnamefont {Yu}}, \bibinfo {author} {\bibfnamefont {J.~T.}\ \bibnamefont {Zhang}}, \bibinfo {author} {\bibfnamefont {N.~R.}\ \bibnamefont {Hutzler}}, \bibinfo {author} {\bibfnamefont {T.}~\bibnamefont {Rosenband}},\ and\ \bibinfo {author} {\bibfnamefont {K.-K.}\ \bibnamefont {Ni}},\ }\href {https://www.science.org/doi/abs/10.1126/science.aar7797} {\bibfield  {journal} {\bibinfo  {journal} {Science}\ }\textbf {\bibinfo {volume} {360}},\ \bibinfo {pages} {900} (\bibinfo {year} {2018})}\BibitemShut {NoStop}%
\bibitem [{\citenamefont {Langin}\ \emph {et~al.}(2021)\citenamefont {Langin}, \citenamefont {Jorapur}, \citenamefont {Zhu}, \citenamefont {Wang},\ and\ \citenamefont {DeMille}}]{Langin2021}%
  \BibitemOpen
  \bibfield  {author} {\bibinfo {author} {\bibfnamefont {T.~K.}\ \bibnamefont {Langin}}, \bibinfo {author} {\bibfnamefont {V.}~\bibnamefont {Jorapur}}, \bibinfo {author} {\bibfnamefont {Y.}~\bibnamefont {Zhu}}, \bibinfo {author} {\bibfnamefont {Q.}~\bibnamefont {Wang}},\ and\ \bibinfo {author} {\bibfnamefont {D.}~\bibnamefont {DeMille}},\ }\href {10.1103/PhysRevLett.127.163201} {\bibfield  {journal} {\bibinfo  {journal} {Phys. Rev. Lett.}\ }\textbf {\bibinfo {volume} {127}} (\bibinfo {year} {2021})}\BibitemShut {NoStop}%
\bibitem [{\citenamefont {Sawaoka}\ \emph {et~al.}(2025)\citenamefont {Sawaoka}, \citenamefont {Nasir}, \citenamefont {Lunstad}, \citenamefont {Li}, \citenamefont {Mango}, \citenamefont {Lasner},\ and\ \citenamefont {Doyle}}]{Sawaoka2025}%
  \BibitemOpen
  \bibfield  {author} {\bibinfo {author} {\bibfnamefont {H.}~\bibnamefont {Sawaoka}}, \bibinfo {author} {\bibfnamefont {A.}~\bibnamefont {Nasir}}, \bibinfo {author} {\bibfnamefont {A.}~\bibnamefont {Lunstad}}, \bibinfo {author} {\bibfnamefont {M.}~\bibnamefont {Li}}, \bibinfo {author} {\bibfnamefont {J.}~\bibnamefont {Mango}}, \bibinfo {author} {\bibfnamefont {Z.~D.}\ \bibnamefont {Lasner}},\ and\ \bibinfo {author} {\bibfnamefont {J.~M.}\ \bibnamefont {Doyle}},\ }\href@noop {} {\  (\bibinfo {year} {2025})},\ \Eprint {https://arxiv.org/abs/2509.01618} {arXiv:2509.01618 [Atomic Physics]} \BibitemShut {NoStop}%
\bibitem [{Note1()}]{Note1}%
  \BibitemOpen
  \bibinfo {note} {For the repumping transitions driven in this work, the light was produced by a Ti:Sapph laser with a wavelengths of $\sim $700-712~nm and intensity of $I\sim 900$~mW/cm$^2$.}\BibitemShut {Stop}%
\bibitem [{Note2()}]{Note2}%
  \BibitemOpen
  \bibinfo {note} {The light is generated by a continuous-wave 899 Coherent Ring dye laser with Rhodamine 640 as the gain medium pumped by 10~W of 532~nm light. When populating states $\protect \tilde {X}^{2}\Sigma ^{+}(12^{0,2}0)$ via cycling on a diagonal line to deplete (transitions (A) and (A') in Figure~\ref {fig:energylevels}), homebuilt ECDLs were used.}\BibitemShut {Stop}%
\bibitem [{\citenamefont {Presunka}\ and\ \citenamefont {Coxon}(1994)}]{SrOHA010}%
  \BibitemOpen
  \bibfield  {author} {\bibinfo {author} {\bibfnamefont {P.~I.}\ \bibnamefont {Presunka}}\ and\ \bibinfo {author} {\bibfnamefont {J.~A.}\ \bibnamefont {Coxon}},\ }\href {https://doi.org/10.1063/1.468171} {\bibfield  {journal} {\bibinfo  {journal} {J. Chem. Phys.}\ }\textbf {\bibinfo {volume} {101}},\ \bibinfo {pages} {201} (\bibinfo {year} {1994})}\BibitemShut {NoStop}%
\bibitem [{\citenamefont {Presunka}\ and\ \citenamefont {Coxon}(1993)}]{SrOH020}%
  \BibitemOpen
  \bibfield  {author} {\bibinfo {author} {\bibfnamefont {P.}~\bibnamefont {Presunka}}\ and\ \bibinfo {author} {\bibfnamefont {J.}~\bibnamefont {Coxon}},\ }\href {https://doi.org/10.1139/v93-211} {\bibfield  {journal} {\bibinfo  {journal} {Can. J. Chem.}\ }\textbf {\bibinfo {volume} {71}},\ \bibinfo {pages} {1689} (\bibinfo {year} {1993})}\BibitemShut {NoStop}%
\bibitem [{\citenamefont {Fletcher}\ \emph {et~al.}(1995)\citenamefont {Fletcher}, \citenamefont {Anderson}, \citenamefont {Barclay},\ and\ \citenamefont {Ziurys}}]{Fletcher95}%
  \BibitemOpen
  \bibfield  {author} {\bibinfo {author} {\bibfnamefont {D.~A.}\ \bibnamefont {Fletcher}}, \bibinfo {author} {\bibfnamefont {M.~A.}\ \bibnamefont {Anderson}}, \bibinfo {author} {\bibfnamefont {W.~L.}\ \bibnamefont {Barclay}},\ and\ \bibinfo {author} {\bibfnamefont {L.~M.}\ \bibnamefont {Ziurys}},\ }\href {https://doi.org/10.1063/1.469482} {\bibfield  {journal} {\bibinfo  {journal} {J. Chem. Phys.}\ }\textbf {\bibinfo {volume} {102}},\ \bibinfo {pages} {4334} (\bibinfo {year} {1995})}\BibitemShut {NoStop}%
\bibitem [{\citenamefont {Presunka}\ and\ \citenamefont {Coxon}(1995)}]{Presunka95}%
  \BibitemOpen
  \bibfield  {author} {\bibinfo {author} {\bibfnamefont {P.~I.}\ \bibnamefont {Presunka}}\ and\ \bibinfo {author} {\bibfnamefont {J.~A.}\ \bibnamefont {Coxon}},\ }\href {https://doi.org/10.1016/0301-0104(94)00330-D} {\bibfield  {journal} {\bibinfo  {journal} {Chem. Phys.}\ }\textbf {\bibinfo {volume} {190}},\ \bibinfo {pages} {97} (\bibinfo {year} {1995})}\BibitemShut {NoStop}%
\bibitem [{\citenamefont {Brazier}\ and\ \citenamefont {Bernath}(1985)}]{Brazier85}%
  \BibitemOpen
  \bibfield  {author} {\bibinfo {author} {\bibfnamefont {C.}~\bibnamefont {Brazier}}\ and\ \bibinfo {author} {\bibfnamefont {P.}~\bibnamefont {Bernath}},\ }\href {https://doi.org/10.1016/0022-2852(85)90345-5} {\bibfield  {journal} {\bibinfo  {journal} {J. Mol. Spectrosc.}\ }\textbf {\bibinfo {volume} {114}},\ \bibinfo {pages} {163} (\bibinfo {year} {1985})}\BibitemShut {NoStop}%
\bibitem [{\citenamefont {Nakagawa}\ \emph {et~al.}(1983)\citenamefont {Nakagawa}, \citenamefont {Wormsbecher},\ and\ \citenamefont {Harris}}]{Nakagawa83}%
  \BibitemOpen
  \bibfield  {author} {\bibinfo {author} {\bibfnamefont {J.}~\bibnamefont {Nakagawa}}, \bibinfo {author} {\bibfnamefont {R.~F.}\ \bibnamefont {Wormsbecher}},\ and\ \bibinfo {author} {\bibfnamefont {D.~O.}\ \bibnamefont {Harris}},\ }\href {https://doi.org/10.1016/0022-2852(83)90336-3} {\bibfield  {journal} {\bibinfo  {journal} {J. Mol. Spectrosc.}\ }\textbf {\bibinfo {volume} {97}},\ \bibinfo {pages} {37} (\bibinfo {year} {1983})}\BibitemShut {NoStop}%
\bibitem [{\citenamefont {Lasner}\ \emph {et~al.}(2022)\citenamefont {Lasner}, \citenamefont {Lunstad}, \citenamefont {Zhang}, \citenamefont {Cheng},\ and\ \citenamefont {Doyle}}]{LasnerVBRs2022}%
  \BibitemOpen
  \bibfield  {author} {\bibinfo {author} {\bibfnamefont {Z.}~\bibnamefont {Lasner}}, \bibinfo {author} {\bibfnamefont {A.}~\bibnamefont {Lunstad}}, \bibinfo {author} {\bibfnamefont {C.}~\bibnamefont {Zhang}}, \bibinfo {author} {\bibfnamefont {L.}~\bibnamefont {Cheng}},\ and\ \bibinfo {author} {\bibfnamefont {J.~M.}\ \bibnamefont {Doyle}},\ }\href {https://doi.org/10.1103/PhysRevA.106.L020801} {\bibfield  {journal} {\bibinfo  {journal} {Phys. Rev. A}\ }\textbf {\bibinfo {volume} {106}},\ \bibinfo {pages} {L020801} (\bibinfo {year} {2022})}\BibitemShut {NoStop}%
\bibitem [{\citenamefont {Hallas}\ \emph {et~al.}(2024)\citenamefont {Hallas}, \citenamefont {Li}, \citenamefont {Vilas}, \citenamefont {Robichaud}, \citenamefont {Anderegg},\ and\ \citenamefont {Doyle}}]{Hallas2024}%
  \BibitemOpen
  \bibfield  {author} {\bibinfo {author} {\bibfnamefont {C.}~\bibnamefont {Hallas}}, \bibinfo {author} {\bibfnamefont {G.~K.}\ \bibnamefont {Li}}, \bibinfo {author} {\bibfnamefont {N.~B.}\ \bibnamefont {Vilas}}, \bibinfo {author} {\bibfnamefont {P.}~\bibnamefont {Robichaud}}, \bibinfo {author} {\bibfnamefont {L.}~\bibnamefont {Anderegg}},\ and\ \bibinfo {author} {\bibfnamefont {J.~M.}\ \bibnamefont {Doyle}},\ }\href@noop {} {\  (\bibinfo {year} {2024})},\ \Eprint {https://arxiv.org/abs/2509.01618} {arXiv:2509.01618 [Atomic Physics]} \BibitemShut {NoStop}%
\bibitem [{\citenamefont {Li}\ \emph {et~al.}(2025)\citenamefont {Li}, \citenamefont {Hallas},\ and\ \citenamefont {Doyle}}]{Li2025}%
  \BibitemOpen
  \bibfield  {author} {\bibinfo {author} {\bibfnamefont {G.~K.}\ \bibnamefont {Li}}, \bibinfo {author} {\bibfnamefont {C.}~\bibnamefont {Hallas}},\ and\ \bibinfo {author} {\bibfnamefont {J.~M.}\ \bibnamefont {Doyle}},\ }\href {http://arxiv.org/abs/2404.03636} {\bibfield  {journal} {\bibinfo  {journal} {New J. Phys.}\ }\textbf {\bibinfo {volume} {27}} (\bibinfo {year} {2025})}\BibitemShut {NoStop}%
\bibitem [{\citenamefont {Kozyryev}\ \emph {et~al.}(2017)\citenamefont {Kozyryev}, \citenamefont {Baum}, \citenamefont {Matsuda}, \citenamefont {Augenbraun}, \citenamefont {Anderegg}, \citenamefont {Sedlack},\ and\ \citenamefont {Doyle}}]{Kozyryev2017}%
  \BibitemOpen
  \bibfield  {author} {\bibinfo {author} {\bibfnamefont {I.}~\bibnamefont {Kozyryev}}, \bibinfo {author} {\bibfnamefont {L.}~\bibnamefont {Baum}}, \bibinfo {author} {\bibfnamefont {K.}~\bibnamefont {Matsuda}}, \bibinfo {author} {\bibfnamefont {B.~L.}\ \bibnamefont {Augenbraun}}, \bibinfo {author} {\bibfnamefont {L.}~\bibnamefont {Anderegg}}, \bibinfo {author} {\bibfnamefont {A.~P.}\ \bibnamefont {Sedlack}},\ and\ \bibinfo {author} {\bibfnamefont {J.~M.}\ \bibnamefont {Doyle}},\ }\href {10.1103/PhysRevLett.118.173201} {\bibfield  {journal} {\bibinfo  {journal} {Phys. Rev. Lett.}\ }\textbf {\bibinfo {volume} {118}} (\bibinfo {year} {2017})}\BibitemShut {NoStop}%
\bibitem [{\citenamefont {Li}\ and\ \citenamefont {Coxon}(1996)}]{CaOHA020}%
  \BibitemOpen
  \bibfield  {author} {\bibinfo {author} {\bibfnamefont {M.}~\bibnamefont {Li}}\ and\ \bibinfo {author} {\bibfnamefont {J.~A.}\ \bibnamefont {Coxon}},\ }\href {https://doi.org/10.1063/1.471762} {\bibfield  {journal} {\bibinfo  {journal} {J. Chem. Phys.}\ }\textbf {\bibinfo {volume} {104}},\ \bibinfo {pages} {4961} (\bibinfo {year} {1996})}\BibitemShut {NoStop}%
\bibitem [{\citenamefont {Ferrari}\ \emph {et~al.}(2003)\citenamefont {Ferrari}, \citenamefont {Cancio}, \citenamefont {Drullinger}, \citenamefont {Giusfredi}, \citenamefont {Poli}, \citenamefont {Prevedelli}, \citenamefont {Toninelli},\ and\ \citenamefont {Tino}}]{Ferrari2003}%
  \BibitemOpen
  \bibfield  {author} {\bibinfo {author} {\bibfnamefont {G.}~\bibnamefont {Ferrari}}, \bibinfo {author} {\bibfnamefont {P.}~\bibnamefont {Cancio}}, \bibinfo {author} {\bibfnamefont {R.}~\bibnamefont {Drullinger}}, \bibinfo {author} {\bibfnamefont {G.}~\bibnamefont {Giusfredi}}, \bibinfo {author} {\bibfnamefont {N.}~\bibnamefont {Poli}}, \bibinfo {author} {\bibfnamefont {M.}~\bibnamefont {Prevedelli}}, \bibinfo {author} {\bibfnamefont {C.}~\bibnamefont {Toninelli}},\ and\ \bibinfo {author} {\bibfnamefont {G.~M.}\ \bibnamefont {Tino}},\ }\href {10.1103/PhysRevLett.91.243002} {\bibfield  {journal} {\bibinfo  {journal} {Phys. Rev. Lett.}\ }\textbf {\bibinfo {volume} {91}} (\bibinfo {year} {2003})}\BibitemShut {NoStop}%
\bibitem [{\citenamefont {Shore}(1990)}]{Shore1990}%
  \BibitemOpen
  \bibfield  {author} {\bibinfo {author} {\bibfnamefont {B.}~\bibnamefont {Shore}},\ }\bibinfo {title} {Analytic rwa solutions with loss},\ in\ \href@noop {} {\emph {\bibinfo {booktitle} {The Theory of Coherent Atomic Excitation}}},\ Vol.~\bibinfo {volume} {1}\ (\bibinfo  {publisher} {John Wiley \& Sons, Inc},\ \bibinfo {year} {1990})\BibitemShut {NoStop}%
\bibitem [{\citenamefont {Williams}\ \emph {et~al.}(2017)\citenamefont {Williams}, \citenamefont {Truppe}, \citenamefont {Hambach}, \citenamefont {Caldwell}, \citenamefont {Fitch}, \citenamefont {Hinds}, \citenamefont {Sauer},\ and\ \citenamefont {Tarbutt}}]{Williams2017}%
  \BibitemOpen
  \bibfield  {author} {\bibinfo {author} {\bibfnamefont {H.~J.}\ \bibnamefont {Williams}}, \bibinfo {author} {\bibfnamefont {S.}~\bibnamefont {Truppe}}, \bibinfo {author} {\bibfnamefont {M.}~\bibnamefont {Hambach}}, \bibinfo {author} {\bibfnamefont {L.}~\bibnamefont {Caldwell}}, \bibinfo {author} {\bibfnamefont {N.~J.}\ \bibnamefont {Fitch}}, \bibinfo {author} {\bibfnamefont {E.~A.}\ \bibnamefont {Hinds}}, \bibinfo {author} {\bibfnamefont {B.~E.}\ \bibnamefont {Sauer}},\ and\ \bibinfo {author} {\bibfnamefont {M.~R.}\ \bibnamefont {Tarbutt}},\ }\href {http://dx.doi.org/10.1088/1367-2630/aa8e52} {\bibfield  {journal} {\bibinfo  {journal} {New J. Phys.}\ }\textbf {\bibinfo {volume} {19}} (\bibinfo {year} {2017})}\BibitemShut {NoStop}%
\end{thebibliography}%

\end{document}